# *Attilio Sacripanti*

# *Uchi Mata Family analysis for coaches and teachers*

Abstract

In this paper it is performed the biomechanical analysis of Uchi Mata family, the unifying vision of biomechanics let coaches and teachers to approach competition and lessons in different way. Uchi Mata and all throws that are in the biomechanical group of Couple applied by trunk – leg showed that these most effective techniques in high level competition , are in fact also energetically less expensive than those of the Lever group. The overview of the sportive judo books that show the different kind of Uchi Mata ( for example) are in term of Biomechanics always the same movement. The Japanese different vision informed also the studies among the world that never analyzed the unifying vision. The complementary tools to increase the effectiveness are also analyzed and few New or Chaotic application of trunk-leg family. The Physical and biomechanical background is analyzed showing that these throws are also bio-mechanically simpler, relying only on closing the distance and Couple application.( GAI + Couple). Their intrinsic simplicity, however, hides a different complexity from those of the Lever which, remember, they need a high motor coordination to be effective. In fact, being able to apply these techniques of Couple require a particular timing ability.

# Attilio Sacripanti

# Uchi Mata Family analysis for coaches and teachers

## 1 Introduction

In the paper of Seoi Family the Japanese and Biomechanical information bring us at very similar and "Classical" solutions. [1]
All the information are well known to Judo coaches and teachers, probably only new way or Chaotic solution were different as content, but ,do not forget, that many Chaotic solutions were introduced by Japanese athletes during competition.
These means that Japanese long experimental researches and Biomechanical point of view are equivalent for the understanding of Seoi Family.
But for Uchi Mata and his Family: all techniques of Couple Group applied by Trunk and Leg, the situation is just a little bit different.
If biomechanics will drive us a different point of view and new knowledge will give us information very different from the "Classical Japanese heritage" that inform the judo world.
Strictly speaking the Biomechanical approach to Uchi Mata and Couple group is very far from the Japanese vision, much more than Seoi family.
Uchi Mata in the agonistic statistics is very close to Seoi about the more utilized technique in high level competitions.
As historical curiosity we can affirm that Uchi Mata it is not into the 200 throws found in the famous Beni Hassan grave in Egypt 3000 B.C. [2]   *Fig 1-6*

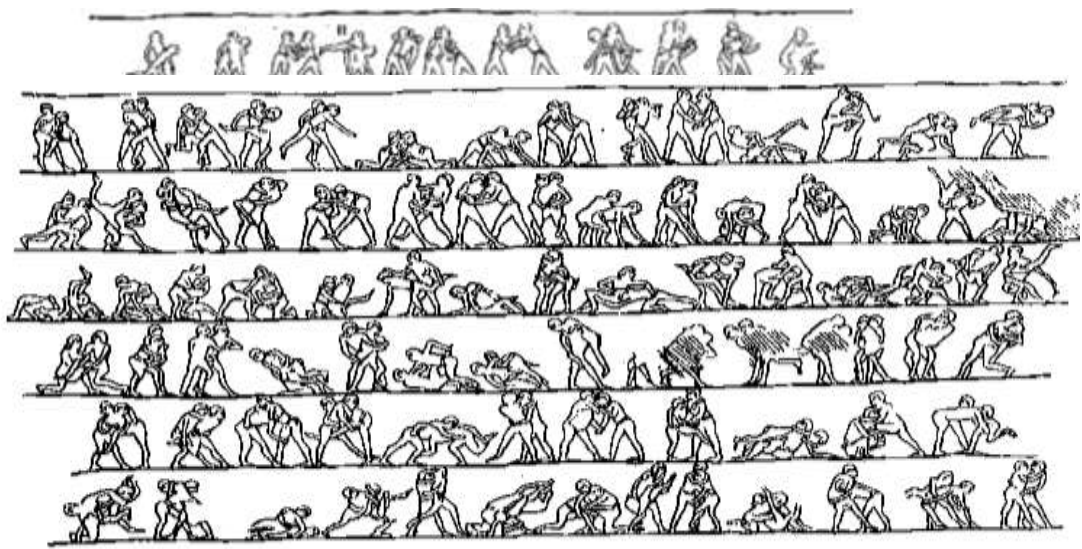

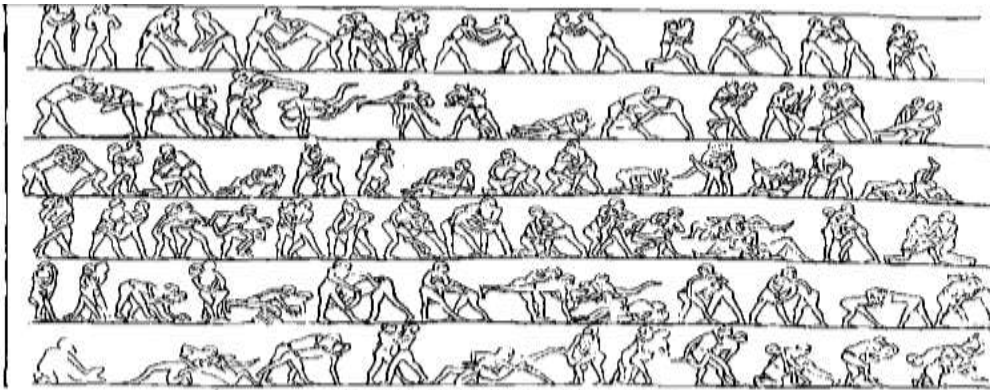

IV. BENI HASAN, TOMB XVII.

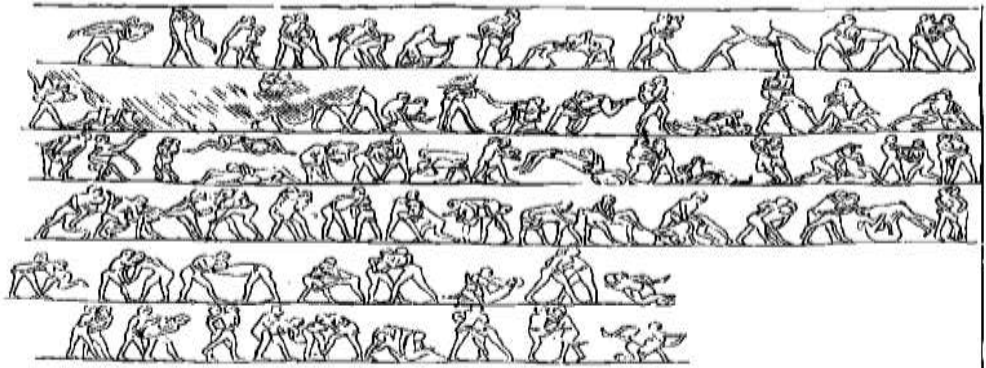

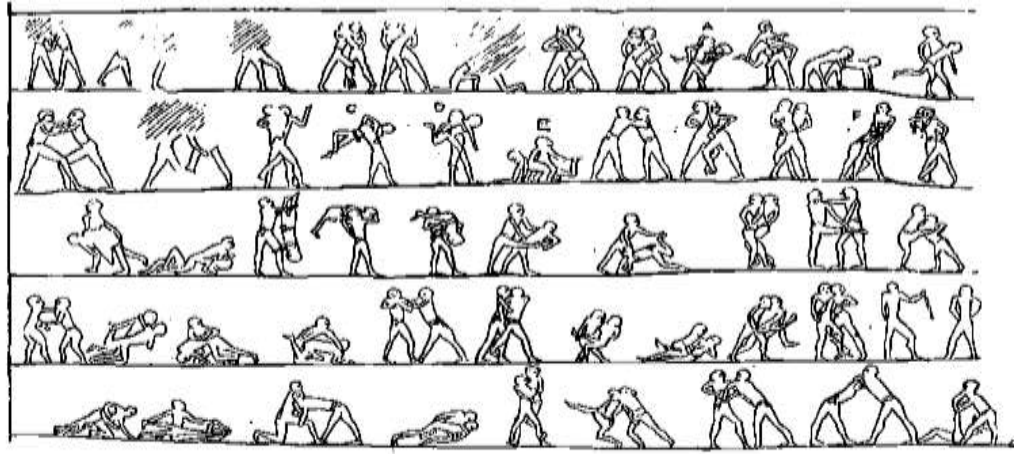

I. BENI HASAN, TOMB XVII. 3000 B.C.

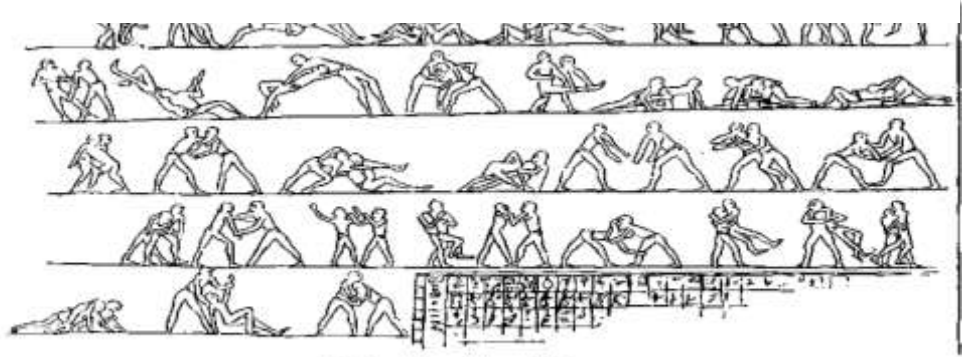

III. BENI HASAN, TOMB XVII. 3000 B.C.

It is strange but ( for courious reader) as couple Throws we can find Kuchiki Taoshi, Kibisu Gaeshi, O Uchi Gari, Ko Uchi Gari, O Soto Gari, but … no Uchi Mata.
Probably the techniques shown in the tomb was also military exercises, and it could be dangerous as military techniques to give his own back to the adversary.

## *2. General Biomechanics of Uchi Mata Family*

However today it is well known that the tools to throw the adversary are two ( Lever : to put a stopping point to rotate adversary around it, Couple: to apply two forces parallel and opposites to rotate the adversary around his Center of Mass) . [3]
The mechanical difference about the final accomplishment is connected to the energy consumption tied to two groups.
For the Lever group ( Seoi, Tai Otoshi, Hiza Guruma) that depends from friction, the techniques need unbalance and the center of mass of Uke shifts in space ( this means positive work and energy consumption).
For the Couple group ( Uchi Mata, O Soto gari, Ashi Arai) Uke turns around his center of mass that can be fixed in space and theoretically these techniques are unbalance and friction independent, and if friction would be zero, for Tori it would be zero work.
Obviously we are in a real world and friction is present however these considerations are made to show why Couple techniques are energetically convenient in comparison to Lever techniques. This information was well known also to Japanese people because already in 1958 some researchers demonstrated the difference in energy consumption among Couple and Lever techniques
For techniques of Couple (trunk leg subgroup) the movement is always based on a support leg while the other leg applies one force the other one's is applied by grips with short or long distance from Uke.
For example in the lateral application ( couple in the Frontal plane) Okuri Ashi Arai as superior force it is sufficient the force applied by arms firmly attached to the trunk, in the front-back application ( couple in the Sagittal plane) Uchi Mata, O Soto Gari, because Uke muscles can apply a stronger defenses than the previous one, it is the whole trunk of Tori connected with grips that applies force.

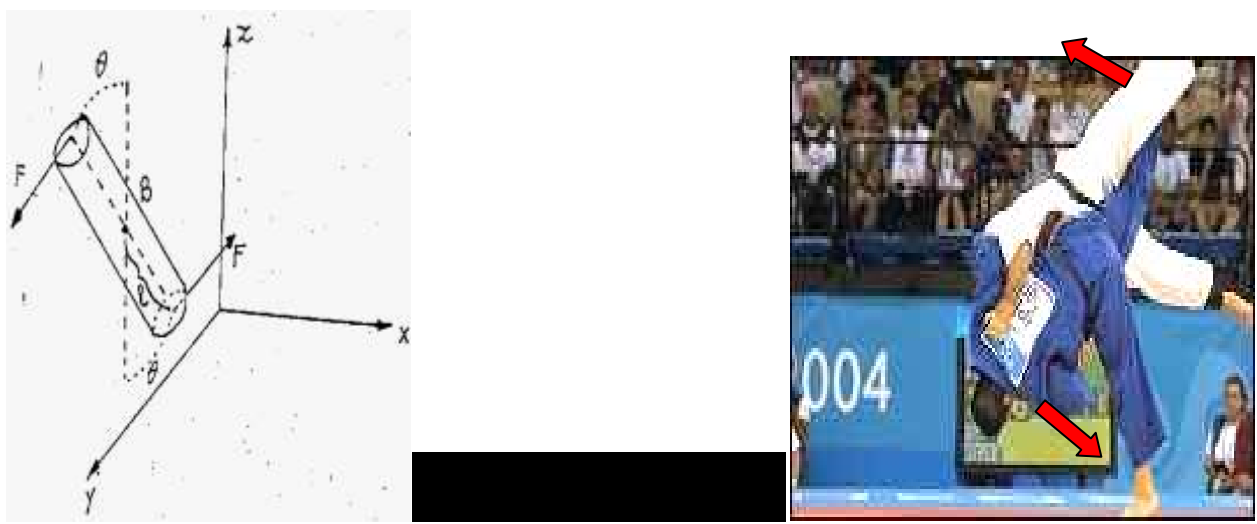

*Fig.1 Basic Mechanics of the Couple techniques (trunk-leg)*

The next table show some results (80) founded by the author that were confirmed in other works 2012 the results are in agreement with worldwide researches [4] [5][6].

Tab. 1

| Groups | Judo -techniques | K Joule |
|---|---|---|
|  | Uchi Mata | 4,2 |
| COUPLE | Ashi Arai | 3,6 |
|  | O Soto Gari | 4,3 |
|  | **Ippon Seoi Nage** | **5,3** |
| **LEVER** | **Koshi Guruma** | **5,8** |
|  | *Tai Otoshi* | *4,9* |

*Tab1 Energy consumption of some judo throws*

### 3. Uchi Mata and Others in books.

Standing Throws classification ( Tachi Waza) was developed by jigoro Kano for the better understanding of inner mechanics of techniques. This classification as well known is grounded on the tools utilized to apply force on Uke's body, namely: arms, hips, legs (Te, Koshi, Ashi). [7] However the proof that the right Uchi Mata mechanics is not well known worldwide is in the different classifications and opinion among also Japanese Masters
For Kano and Kodokan Uchi Mata is Ashi Waza, same classification for Mifune, Okano, and Daigo [8] [9] [10] while for Kazuzo Kudo, Kawaishi e Nakano it is Koshi Waza, [11] [12] in Europe there are also more differences : for Koizumi it is Tsumazukase Waza ( reaping techniques),[13] for Geesink it is a Tachi Waza with moving leg, [14] for Gleeson lifting techniques ( subgroup rotation techniques) [15] for Dr, Andres Kolychkine Thomson Cuban Master, Uchi Mata is a mixed technique of hip and leg.[16]
Really speaking these differences are nonsense the mechanics of techniques is very clear a Couple produced by trunk and leg in the Sagittal Plane of symmetry , dynamically speaking if the contact is with leg or hip is not essential, the Couple is always produced by trunk and leg that theoretically would lie on the same line; then Classic Uchi Mata Tori's right leg mows Uke's left leg, Hip or Koshi Uchi Mata the leg mows in the center , and Okurikomi Uchi Mata the right legs mows the right leg of Uke, however biomechanically speaking there is the same movement and the same technique. Fig.2
The simple biomechanics of these techniques (appropriate shortening distance and Couple tool application) let us able to understand that all the different techniques of Uchi Mata classified in the textbooks as ***different techniques*** biomechanically speaking are only different way to shorten distance and a monotonous Couple application.

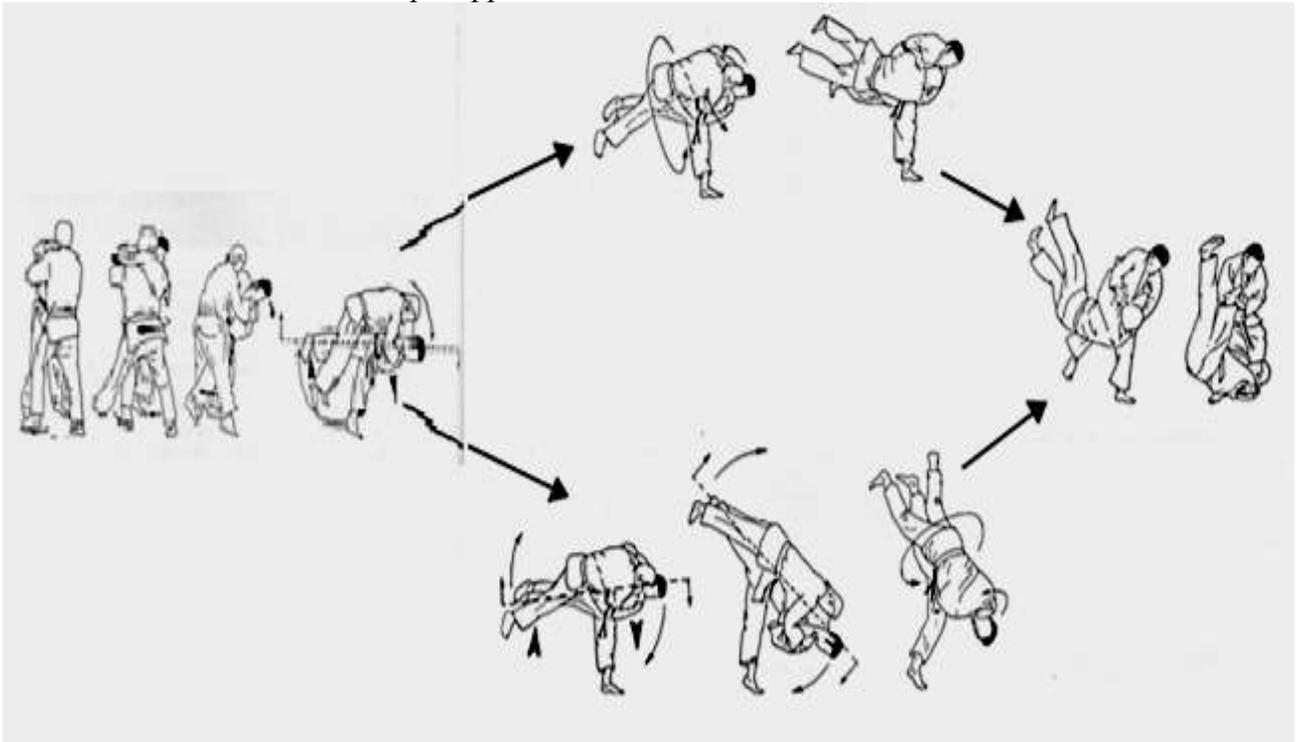

*Fig.2 Same mechanics for two "differents" Uchi Mata.* **[17]**

In the next pictures from the text of Sato and Okano there are show three type of Uchi Mata in which it is clear that the throwing tool is always the same : Couple applied by trunk and leg the differences are in the relative bodies positions and mowing leg contact point.
There are no differences, from the theoretical and biomechanics point of view, however the differentiation of contact points and trajectories to shorten distances between athletes in practical way change only the energetic of throw actions, and the so called Koshi Uchi Mata seems to be the most expensive among the three.
Anyway these three different variation of Uchi Mata (always the same Biomechanics movement) are less expensive of Lever techniques.
We can see Uchi Mata Classic, Okurikomi Uchi Mata, and Koshi Uchi Mata, performed by Sato from the book Vital Judo .Fig 3-16.

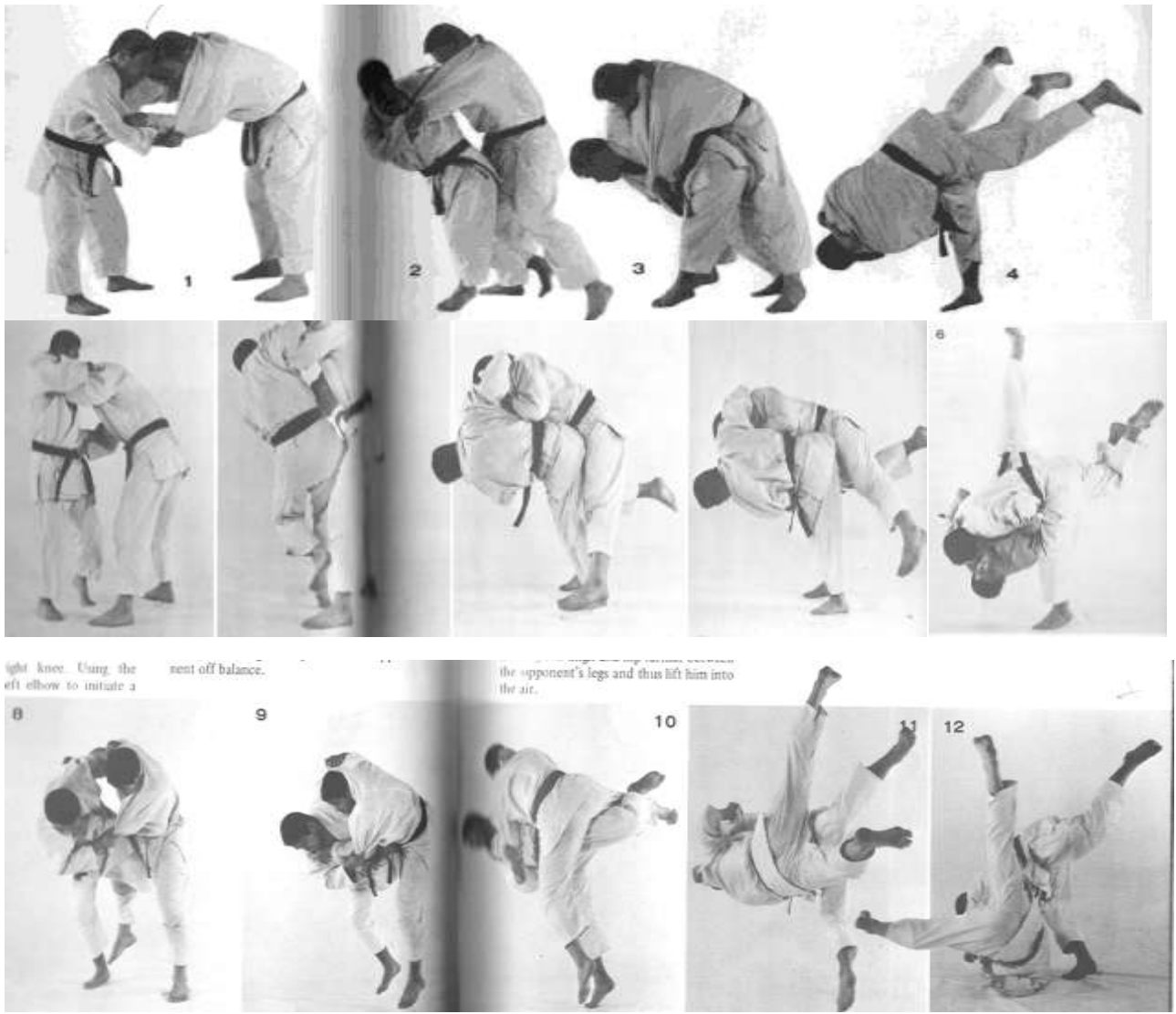

*Fig.3-16 Vital Judo different way to apply Couple tool by Sato [9]*

From these pictures it is clear the mechanical simplicity of these Couple throws: GAI + Couple, very different from Lever: GAI + SSAI + ISAI + Lever.
How many are the trunk-leg throws?
If we analyze the human body's symmetry planes into the sagittal plane there are two classical way to apply couple tool Uchi Mata with a rotational approaching trajectory ( GAI ) of 180° and O Soto Gari with a right approaching trajectory (GAI).
From the classical Japanese point of view the following figures are different techniques but for biomechanics they are always the same tool applied. Fig.17-18

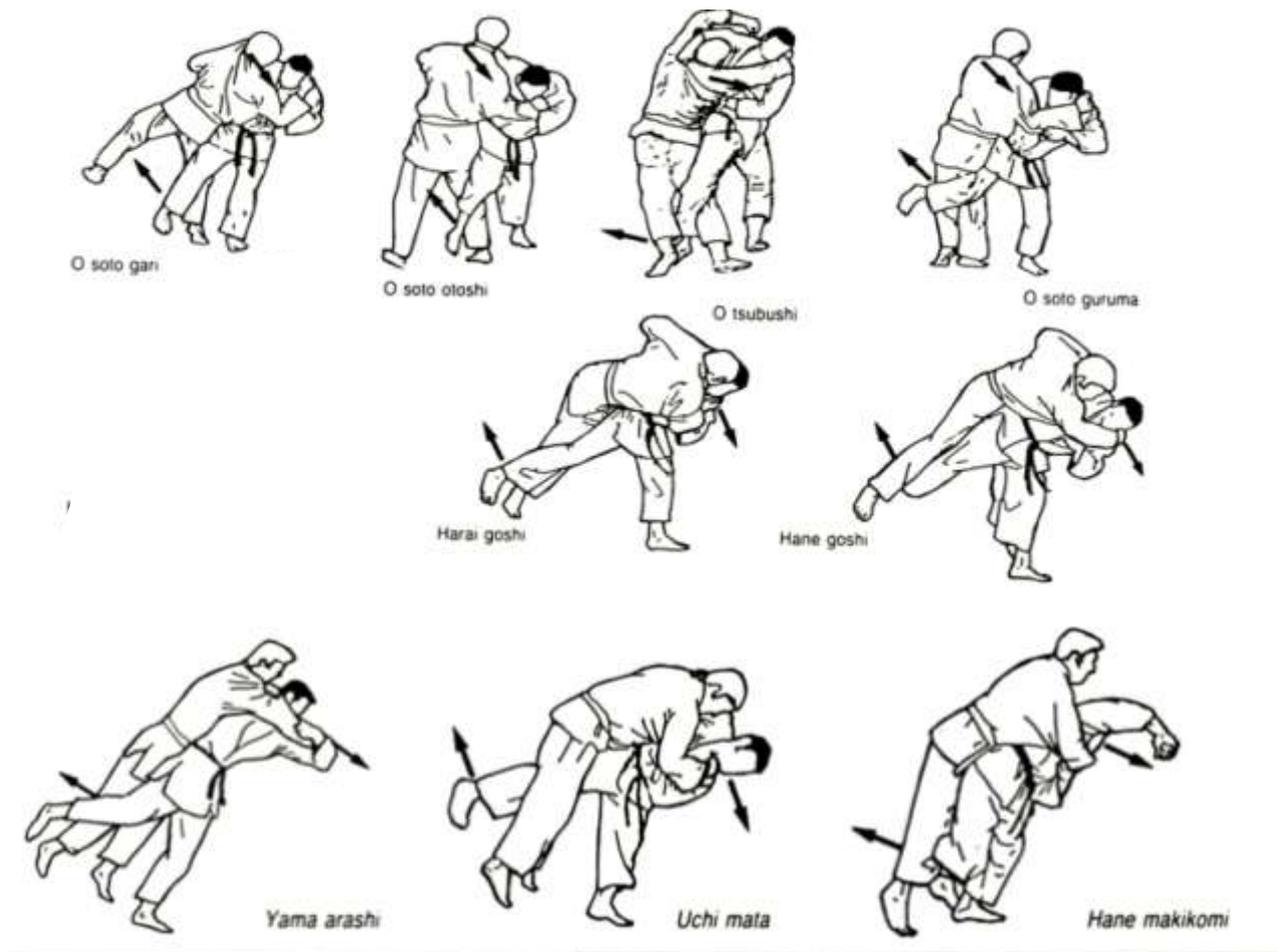

*Fig.17-18 Classical vision trunk-leg techniques. [18]*

The biomechanical point of view shows all unifying capability, when considering Couple as tool.
In this situation classical Japanese view in which there are a lot of different techniques like Uchi Mata, O Soto Gari, Harai Goshi is wiped away and all throws (for example applied by trunk and leg ) are the application of the same tool with different application angles and placement trajectories.
Then athletes must shorten the distance among them (GAI) always 180° of rotation or straight line and apply Couple with one support leg, whatever relative position they have (in some condition also without unbalance).
In this point of view for Tori considering only the throwing action Uchi Mata and O Soto Gari are the same Couple application differs only the asymmetry of Uke's body (front, back) .

## *4. Worldwide researches on Uchi Mata family*

Worldwide study on Uchi Mata family ( Couple Group throws by Trunk Leg) are numerous as studies on Seoi Family, but the main difference is that the Japanese heritage in Couple techniques was so heavy that many work were too focalized in small particular without a powerful common vision like Biomechanical point of view.
Despite this handicap many data obtained (especially for Uchi Mata) are very interesting.
The specific focalization of the researchers on Uchi Mata is understandable because statistical frequency and effectiveness of this throw in competition is high than the Seoi, as it is possible to see from the following tables. tabb2, 3

| | FRA | | JPN | | URSS | | Autres | |
|---|---|---|---|---|---|---|---|---|
| | Uchi-mata | 25,5 % | Uchi-mata | 15,8 % | Uchi-mata | 11,4 % | Suwari-seoi-nage | 13,8 % |
| | O-uchi-gari | 11 % | Suwari-seoi-nage | 13,3 % | Seoi-nage, Kata-guruma | 9,6 % | Uchi-mata | 13,4 % |
| | O-soto-gari | 7,7 % | Ko-uchi-gari | 10,7 % | Suwari-seoi-nage | 7 % | O-uchi-gari | 8,7 % |
| | Sode-tsuri-komi-goshi | 7,4 % | O-uchi-gari | 9 % | Kuchiki-daoshi | 7,7 % | Ko-uchi-gari | 7,4 % |
| | Ko-uchi-gari | 7,4 % | O-soto-gari | 7,3 % | O-soto-gari | 7,4 % | Seoi-nage, Kata-guruma | 7,4 % |
| | Kuchiki-daoshi | 7,4 % | Tomoe-nage | 6 % | Sode-tsuri-komi-goshi | 7 % | Hara-goshi | 7 % |
| | Suwari-seoi-nage | 5,3 % | Seoi-nage, Kata-guruma | 5,1 % | Tai-otoshi | 7 % | | |

*Tab. 2 % of throws utilization of French, Japanese and Russian Athletes. [19]*

| | FRA | JPN | URSS | Autres |
|---|---|---|---|---|
| 1 | Uchi-mata | Suwari-seoi-nage | Seoi-nage, Kata-guruma | O-soto-gari |
| 2 | Kuchiki-daoshi | Seoi-nage, Kata-guruma | Uchi-mata | Suwari-seoi-nage |
| 3 | Seoi-nage, Kata-guruma | Sode-tsuri-komi-goshi | Ura-nage | Sode-suri-komi-goshi |
| 4 | O-uchi-gari | Ko-soto-gari-gaké | O-uchi-gari | O-uchi-gari |
| 5 | O-soto-gari | Tomoe-nage | Sode-tsuri-komi-goshi | Tai-otoshi |
| 6 | Hiza, Sasae | O-soto-gari | Maki-komi | Tomoe-nage |

*Tab. 3 Throws effectiveness of French, Japanese and Russian Athletes. [19]*

If we see at conditioning and specific muscular training it is very important the knowledge of the muscular structures interested in the movement and the specific intervention evolution
during throw execution. in time.
In the next figures there are shown the muscles that perform dynamical and statics contractions
However we must remember that the movement is always the same with small variation of the
connected to different kind of grips. Fig. 19-20

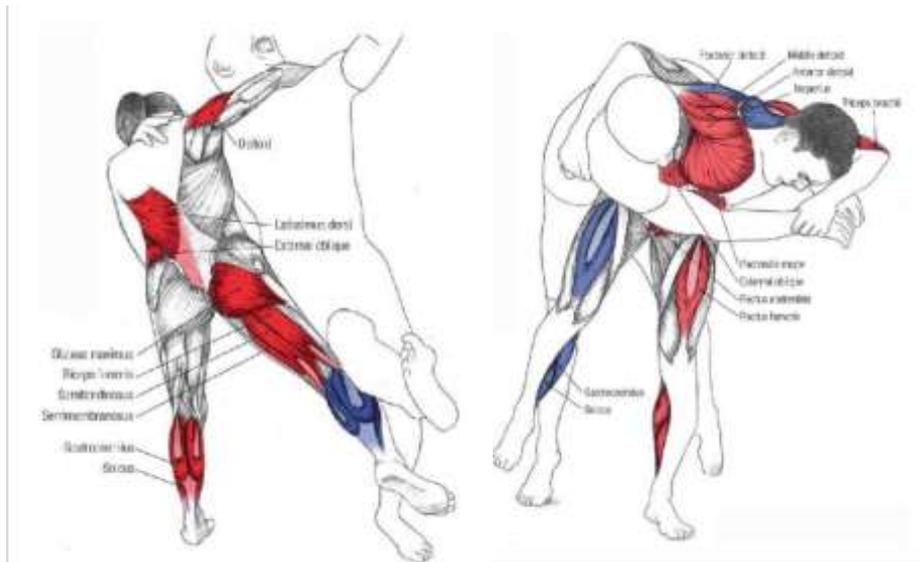
*Fig.19-20 muscular activity during Couple Trunk .Leg throw application
( static red,and dynamic blu ) [20]*

The true effectiveness of Uchi Mata in high level competitions it is the reason that a lot of studies were performed about it.
It will follow an overview of worldwide researches, to show how many different point of view and how many technical aspects are underlined into the world laboratories.
However researches so focalized loose the unifying Biomechanical vision on the matter.
Couple techniques can be grouped also in function of trajectory to gain contact (GAI) :
[ a) right, b) invard rotation180°, c) rotation di 90°]. [21]
A practical vantage from the biomechanical vision is that singling out the basic movement of techniques it is possible to improve technical training by machines.
In the following figures there are shown two specific ergometers for Lever and Couple techniques at Poitiers University France. fig.21.

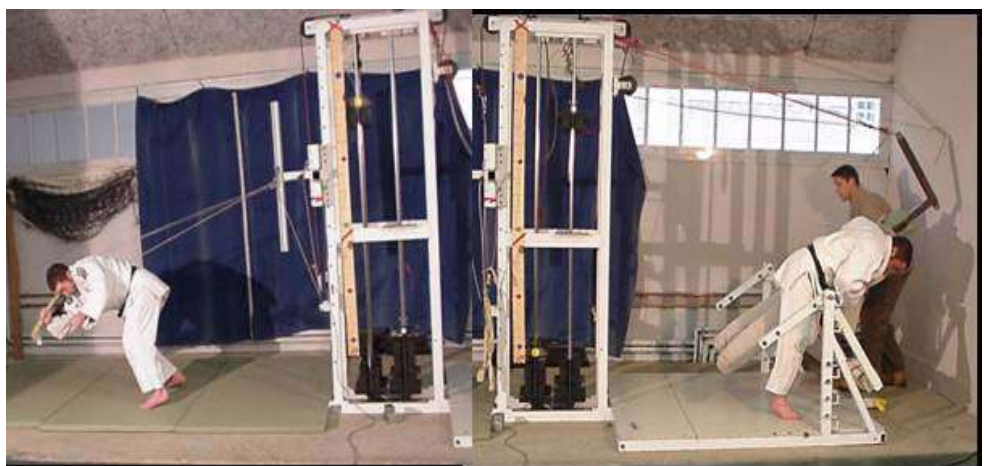
*Fig 21 French Ergometers based on basic actions (Lever [Seoi ] and Couple [Uchi Mata])[22]*

In France some very deep studies were developed about techniques of Couple with inward rotation
 ( Uchi Mata).

Studies performed with high speed camera and movement analysis were able to single out the Action Invariant ( equal movements) of techniques, in connection with their kinetic and kinematic characteristics ; in the next figures it is shown a crono-photogrammetry of Uchi Mata held in Poitiers University.

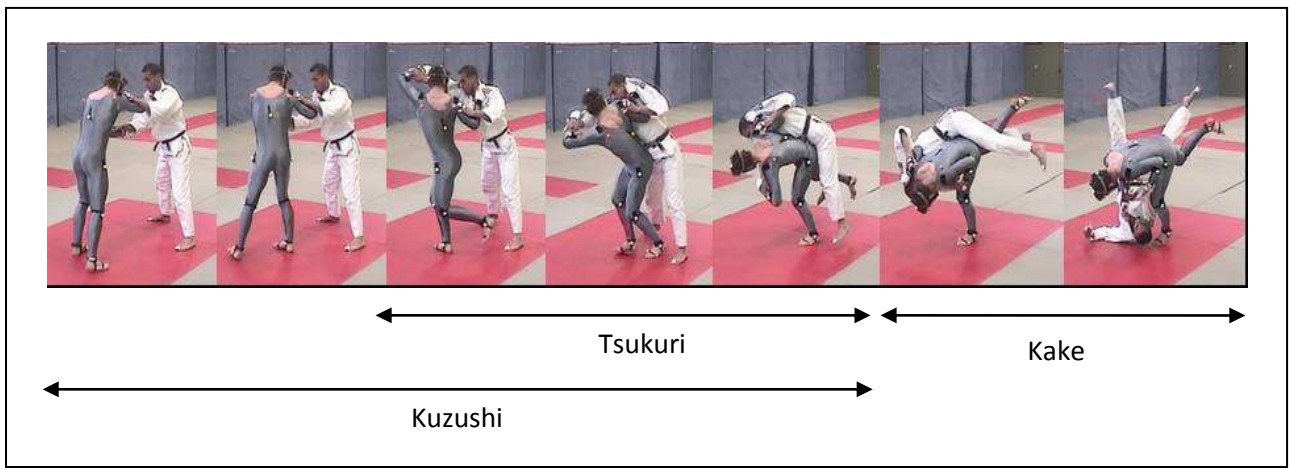

*Fig.22 Crono-photogrammetry of Uchi Mata ( France)[23]*

During these years Japan performed a lot of researches focalized on Uchi Mata , starting from the far 1958 for the Kodokan Association for the Scientific Studies on Judo , in which were studied kinetic of Center of Mass [24], Energetic of throws [25], muscles electromyography showed the most active during throw, these old studies can be associated to the data of figures 19-20 [26], till to 1969 that were studied the regulation of respiration during the throwing actions. F ig 23 [27]

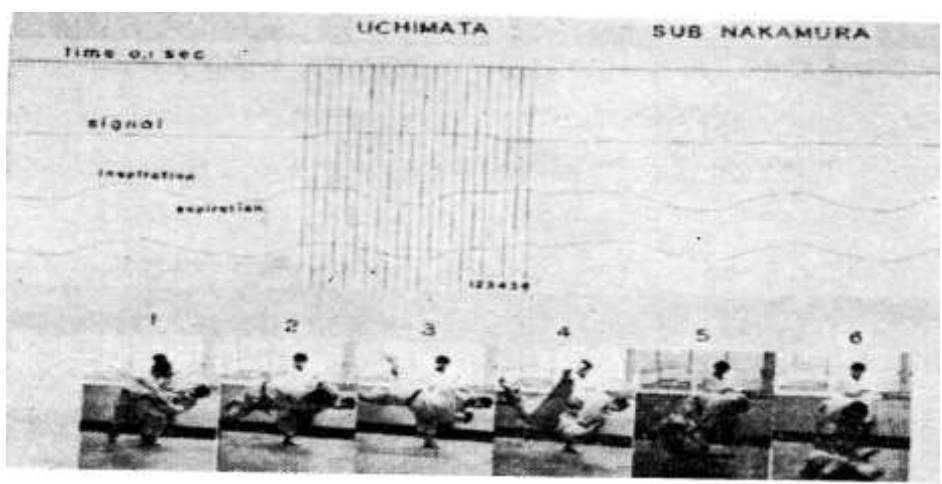

*Fig 23 Nakamura's Uchi Mata [27]*

Comparative evaluation between Tori's CM motions in Classical Uchi Mata and Flamingo style throw was presented in 1988 in Montana in the same congress I presented a work on free style wrestling.. Figg. 24-25

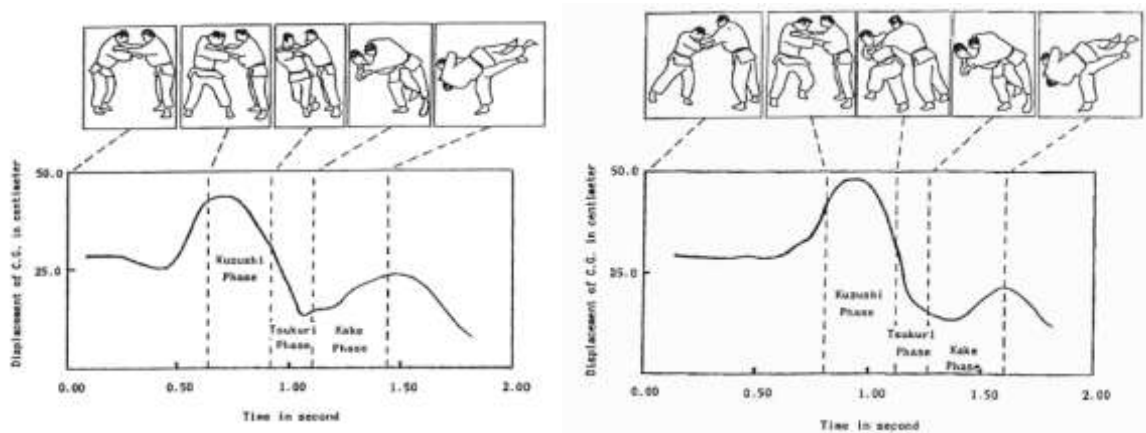

*Fig 24-25 Comparative motion of Center of Mass in UchiMata ( Classic vs Flamingo) [28]*

In modern time the complete biomechanical characterization of Uchi Mata with the set complete of kinetic and kinematic quantities has been performed in the Japanese Universities.

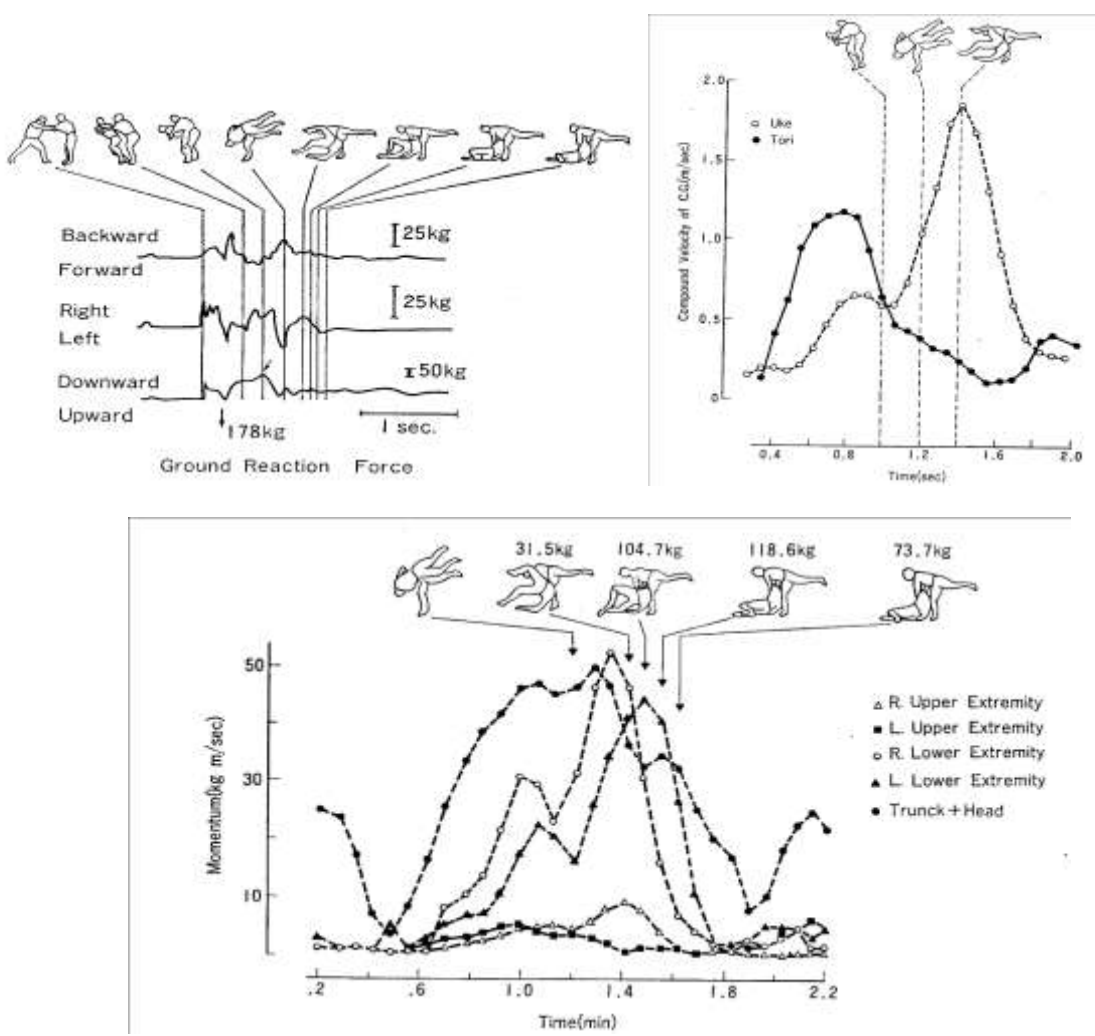

*Fig 26-28 Uchi Mata deep Japanese studies [29]*

In the West interesting studies were performer in US by Prof. Rodney Imamura who collaborated also with Kodokan , in the following figures there are shown results about Harai Goshi

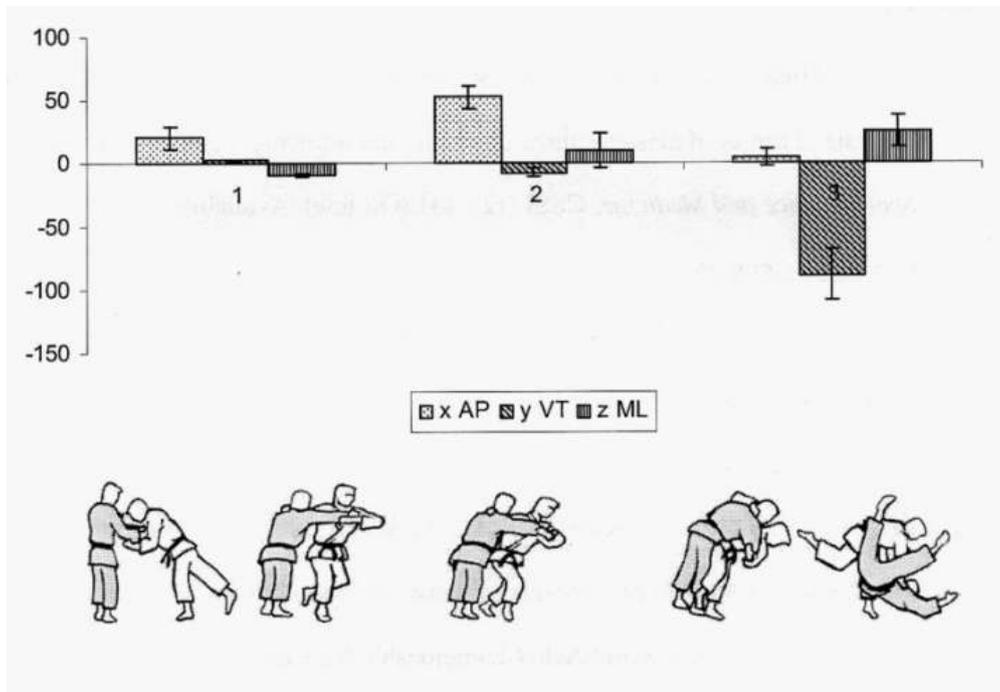

*Fig.29 Harai Goshi US studies [30]*

In Germany Manfredi Vieden analyzed the same techniques from another point of view : study of the momentum transferred during the collision to Uke in comparison with Hane Goshi in the following figures some results.

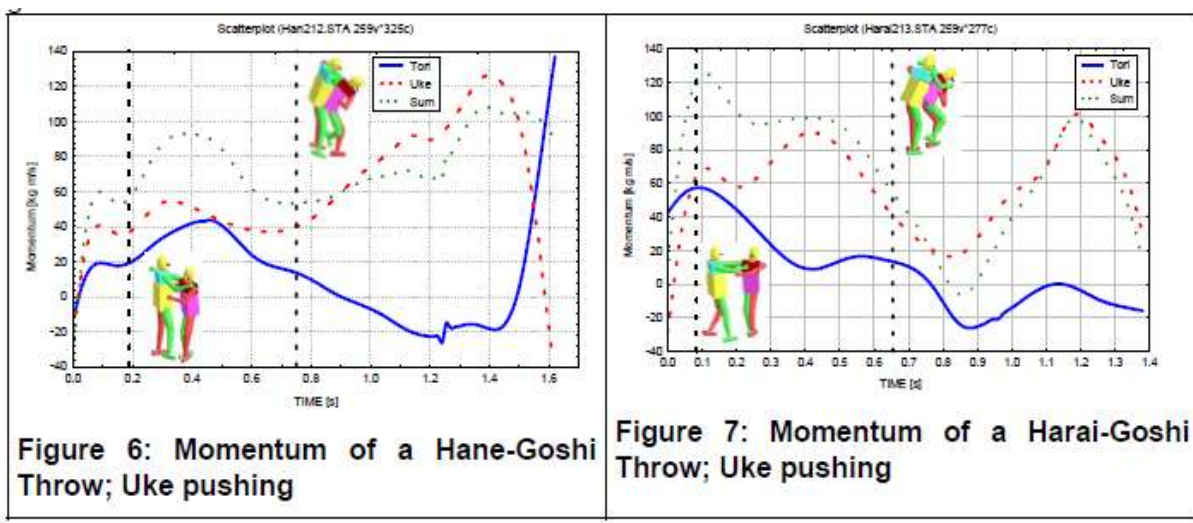

*Fig 30-31 German studies on comparison of trunk leg techniques [31]*

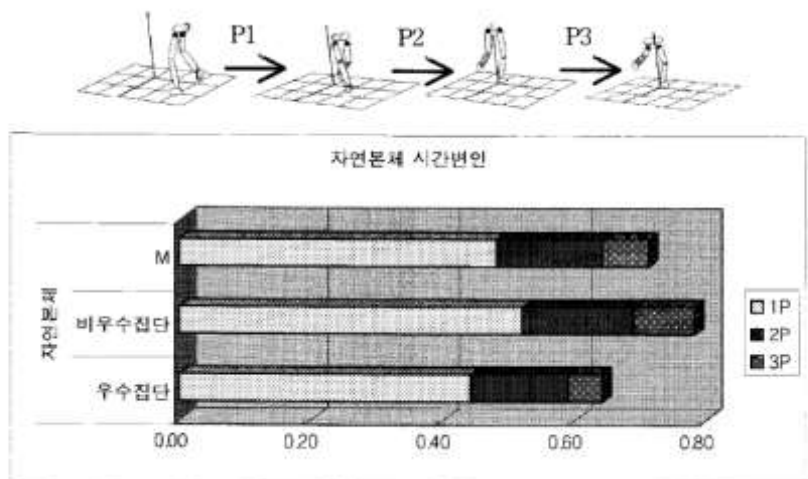

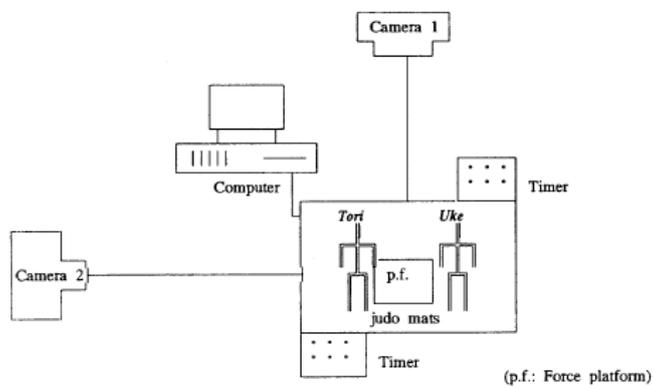

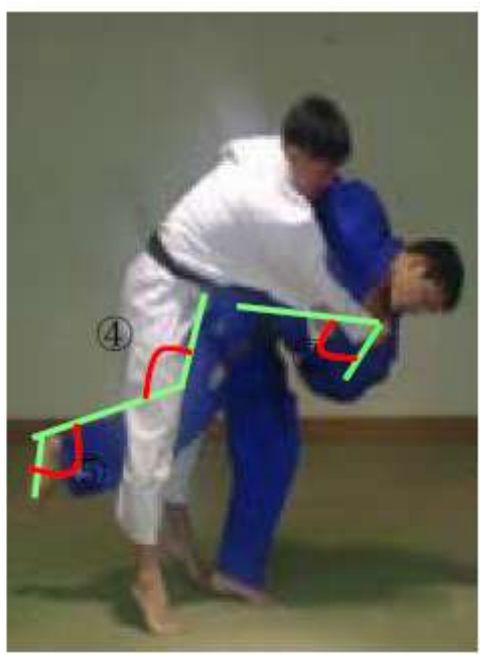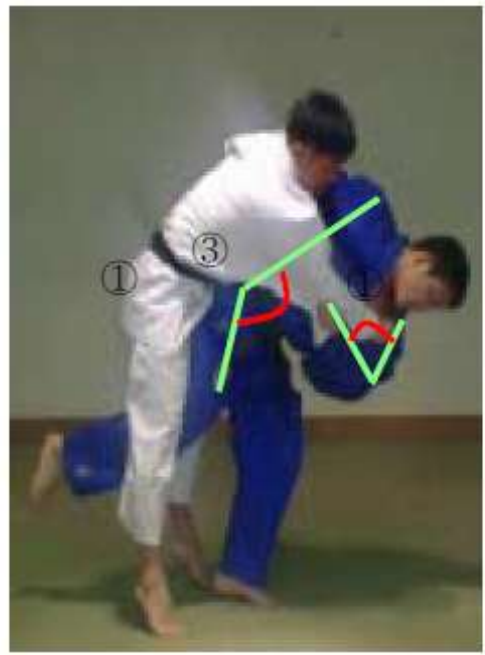

*Fig 34-36   Korean Studies on Uchi Mata[32][33][34]*

Uchi Mata throws considering the worldwide application in high level competitions was studied not only as specific research but also as University Thesis in many countries, as example they are shown the results of a very big Spanish Doctoral Dissertation, in which there are analyzed not only the static parameters but also the most complex dynamic situations developed during competition.

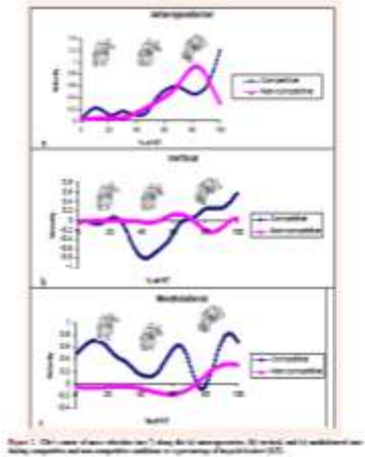
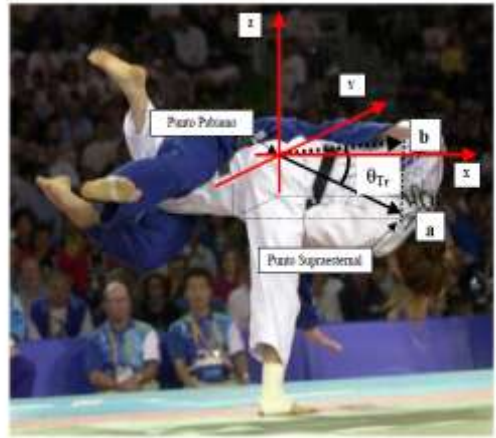

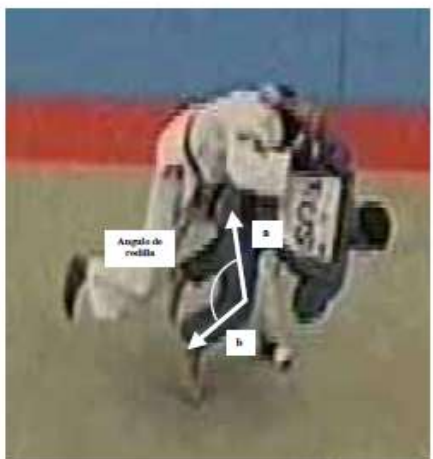
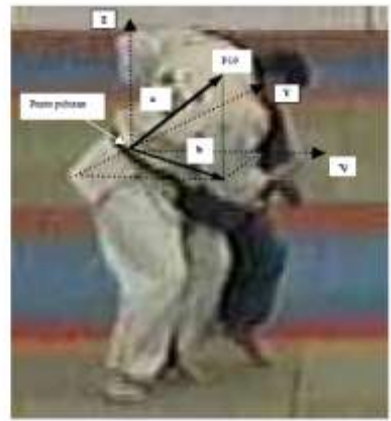

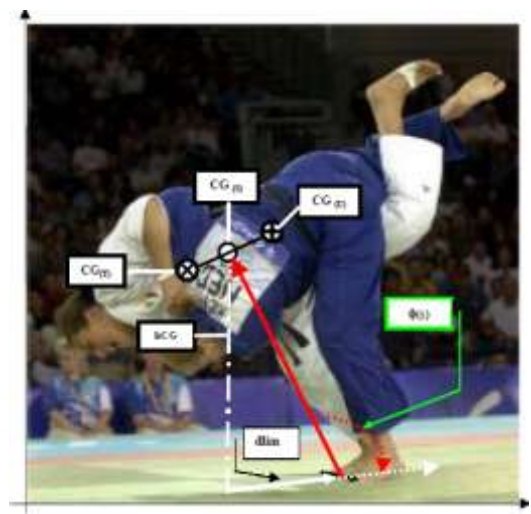

*Fig 32-36 Spanish results for Uchi Mata thesis. [35]*

## 5. *Direct Attack: complementary tactical tools.*

In high level competitions the most important parameter for specialist of Uchi Mata family is attack *Timing* when techniques are applied in Couple of Athletes moving situation.
Timing for these techniques is essential to achieve the better positioning to apply Couple Tool.
However the increased defensive and acrobatic capabilities of worldwide champions make, also with good timing, very difficult to achieve the best relative position to apply Couple.
Then to solve these situations athletes perform tactical complementary tools, that enhance effectiveness of throwing techniques applied by direct attack.
The Uchi Mata Family (Group of Couple applied by Trunk-leg) has many possibilities to enhance his own effectiveness by these complementary tools.
For knowledge follow the most applied complementary tools of Couple trunk-leg applied in high level competition considering the symmetry plane of Human Body.
**Rotational Application**
**Sideway Application**
**The Couple applied in the Sagittal Plane could be most effective if it is followed by a Couple applied in the transverse plane.**
**The Couple applied in the Sagittal Plane could be most effective if it is followed by a continuous torque in the same plane.**
**The Couple applied in the Sagittal Plane could be most effective if it is applied changing time of forces like lever.**
It is well known that rotational movement are already present in the final part of some techniques of Couple group, then it is useful to study some rotational variations that could be utilized in competition to enhance throws effectiveness.
How it is possible to find the right complementary tool?
- Considering the adversary's defensive capability!

Analyzing the human body structure, this means to find the right direction in which human body muscles are less able to resist to the throwing action.
To understand that means to identify trajectories of better use of Energy, which are connected to the shifting's paths of adversary's Center of Mass in space.
Considering the Couple mechanics it is clear how to enhance techniques.
All couple techniques increase their effectiveness or with rotational variations or with very fast changing of their specific mechanics applying the Couple forces not simultaneously but in two steps, or modifying the attack direction, or applying in a row subsequent rotations.
For example applying Couple in specific Adversary bodies diagonal directions: *Fig 37-40*

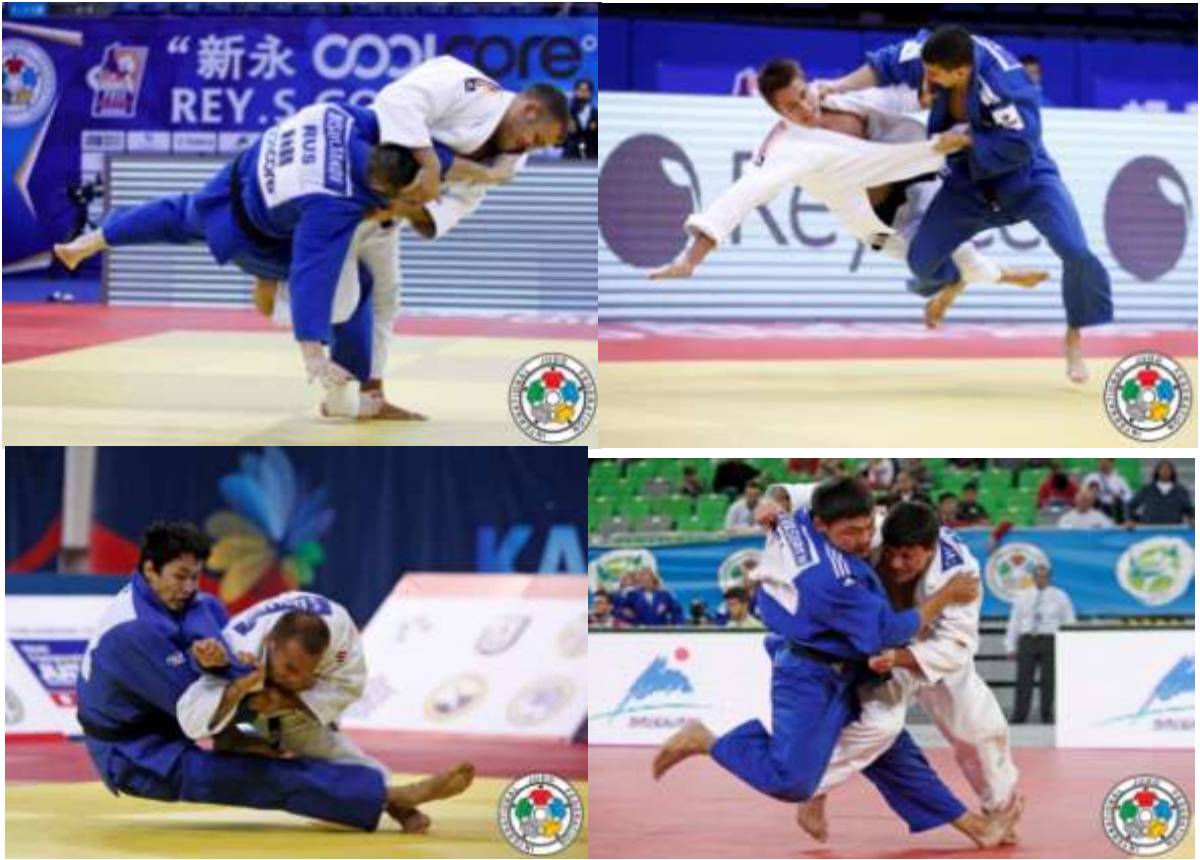

*Fig 37-40 Diagonal applications of Innovative throws belonging to Couple group : O Soto Gari, Ko Uchi gari, and a new Uchi Mata henka (?). [36]*

Another method is to connect at a rotation in the Sagittal plan (Vertical) a following rotation in the Transverse plan (Horizontal) to enhance the effectiveness specifically for the subgroup Trunk-leg. In this group we have Uchi Mata and O Soto Gari; that biomechanically speaking are for Tori the same way to apply the Couple see: *Fig 41*; although in the Japanese classical vision they are two different movements and techniques.

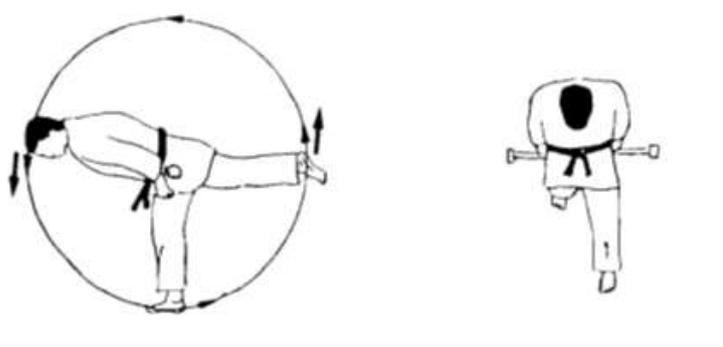

*Fig.41 Trunk-leg same movement to apply Couple in O Soto Gari and Uchi Mata the different vision (two techniques) borns from difference front /back of Uke. [18]*

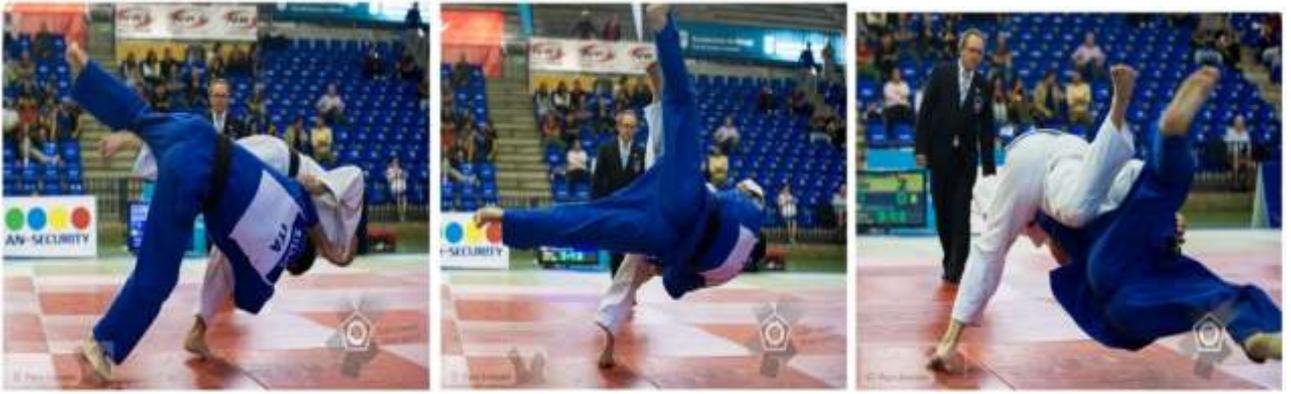
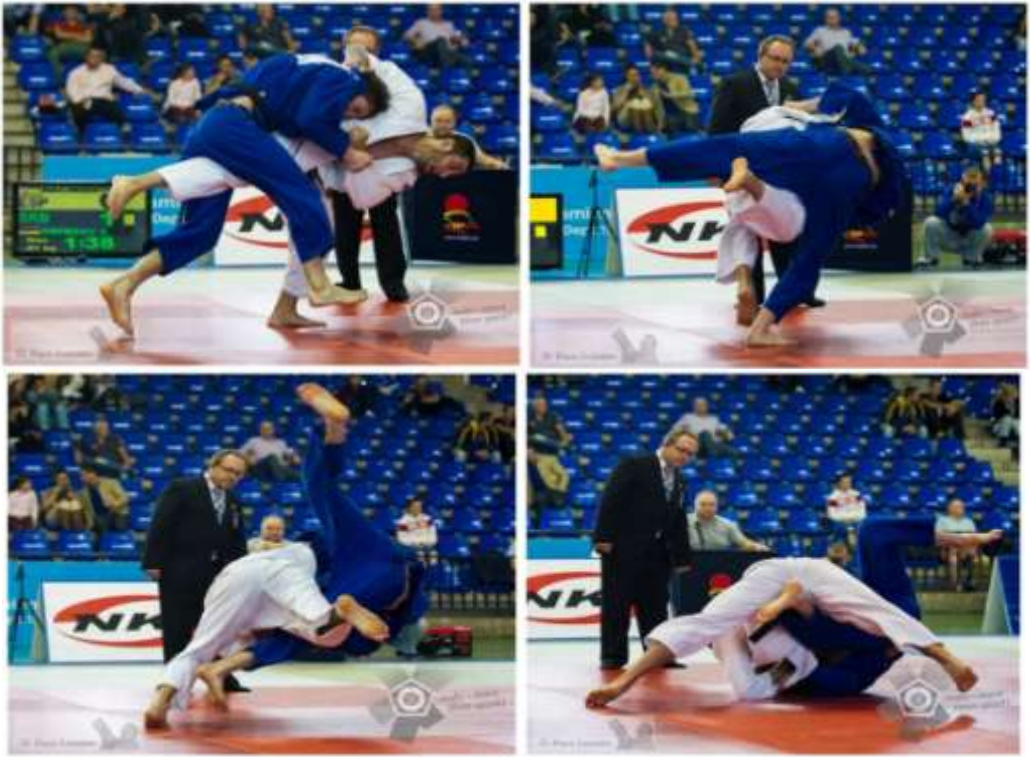

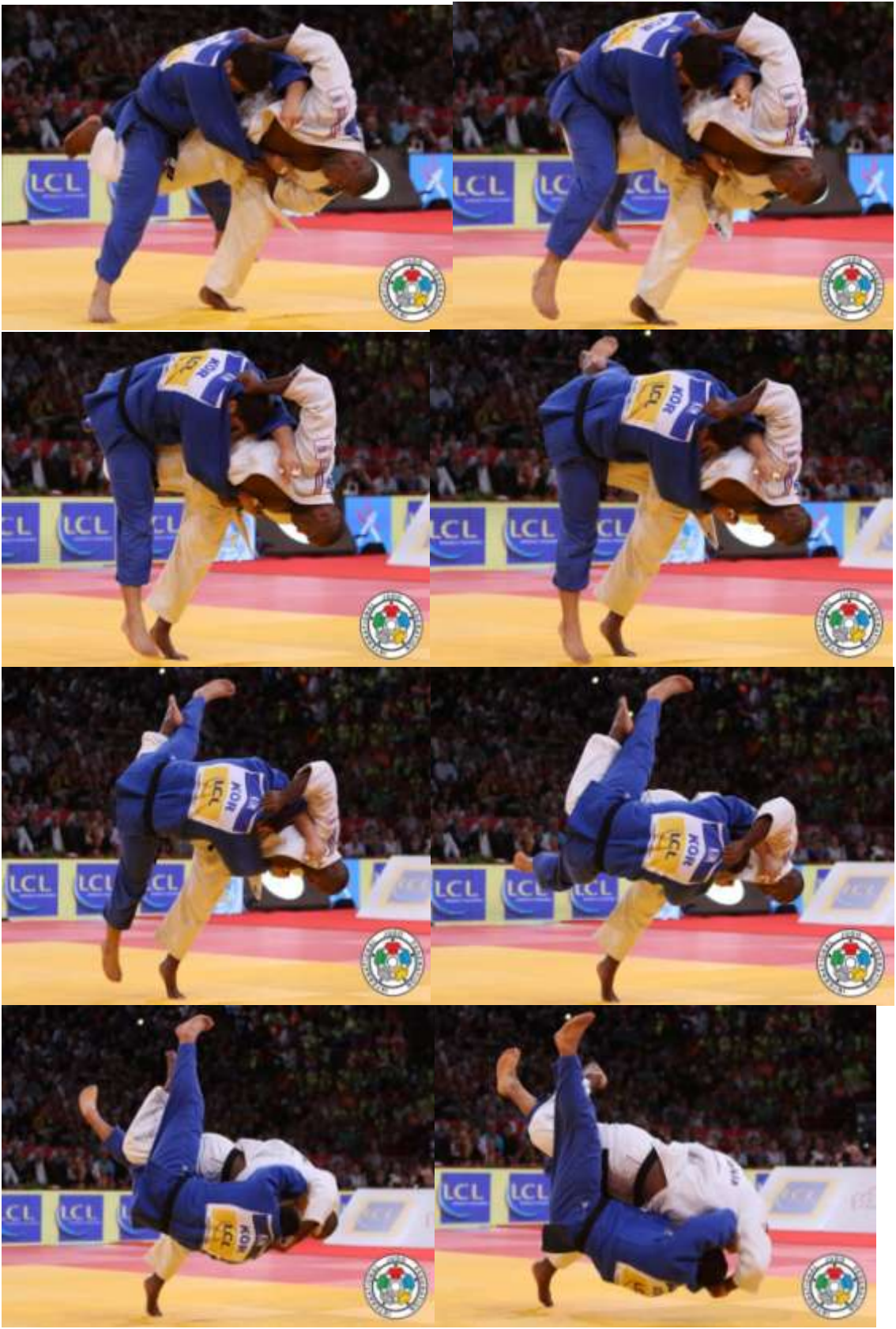

*Fig 42-56 three different methods of rotation in the Transverse plan applied by Tori to help Uchi Mata throwing action. Biomechanics suggest that the same complementary tool can be applied on O Soto Gari. [36]*

The High defensive skill in high level competition makes the perfect Couple application very hard, the third way to enhance effectiveness of the Trunk-Leg throws is to drive Uke Center of Mass along a pseudo-circular trajectory ended it by Mawarikomi , which applies a lever component that is able to throw on his back Uke.
See: the following figure 57.

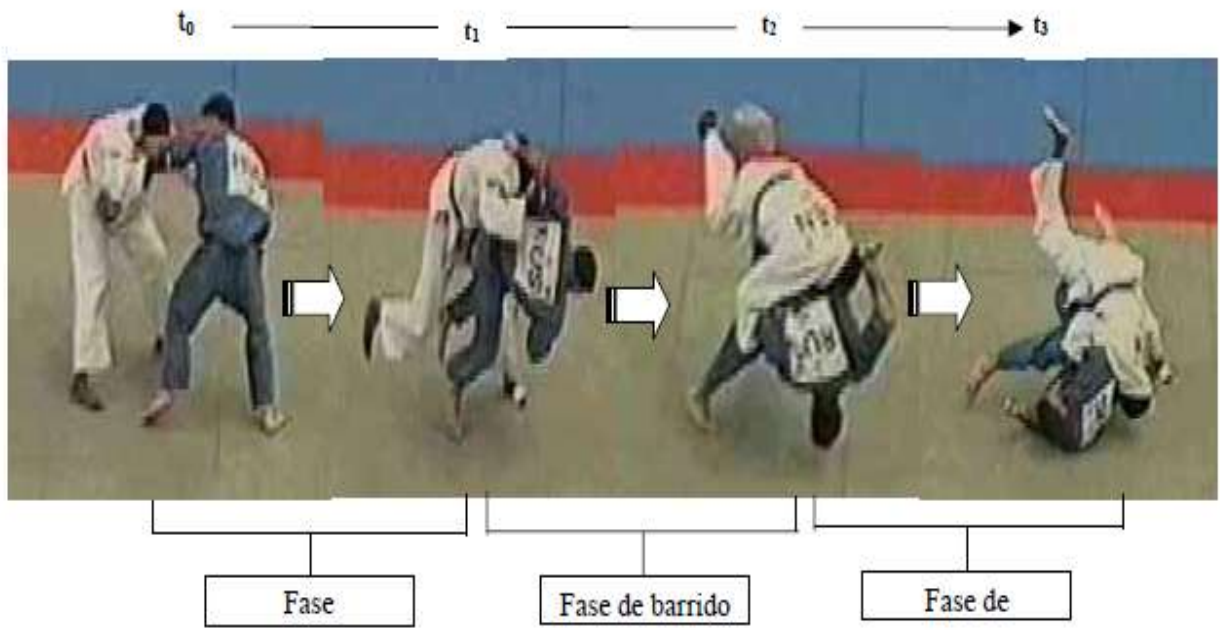

*Fig 57 Mawarikomi in Uchi Mata ( Spanish Thesis)[35]*

If we remember the inner mechanic of Couple Group the Uke's Center of Mass is ideally fixed in space, while translation of center of Mass in space is a characteristic of Lever Group all this implies that in some characteristic situation enhancing Couple techniques effectiveness means to change Couple in Lever, or in physical word to apply a momentum of Force than a couple of force.
Then the enhancement of effectiveness of judo throwing techniques in dynamic situation during high level competition is, among others complementary tools, also grounded on the chance to change continuously Couple in Lever and vice versa.
In Biomechanical terms a true champion will understand when to change the mechanics of his throw to win!
Biomechanics let us to understand clearly how it can happen, for example for Couple Throw.
Biomechanics clarifies that this would be possible in two ways:

1) The mechanics of a Couple throw can be changed in Lever delaying the application of one force. In fact the simultaneous application produces a Couple and consequently a torque, the application at different time produces a momentum and consequently a torque, with different energy consumption.
   Then the application of a momentum moves the Uke's Center of Mass along a pseudo-circular trajectory increasing the effectiveness of Couple Throws in some specific situations.

. *Fig 58,63*

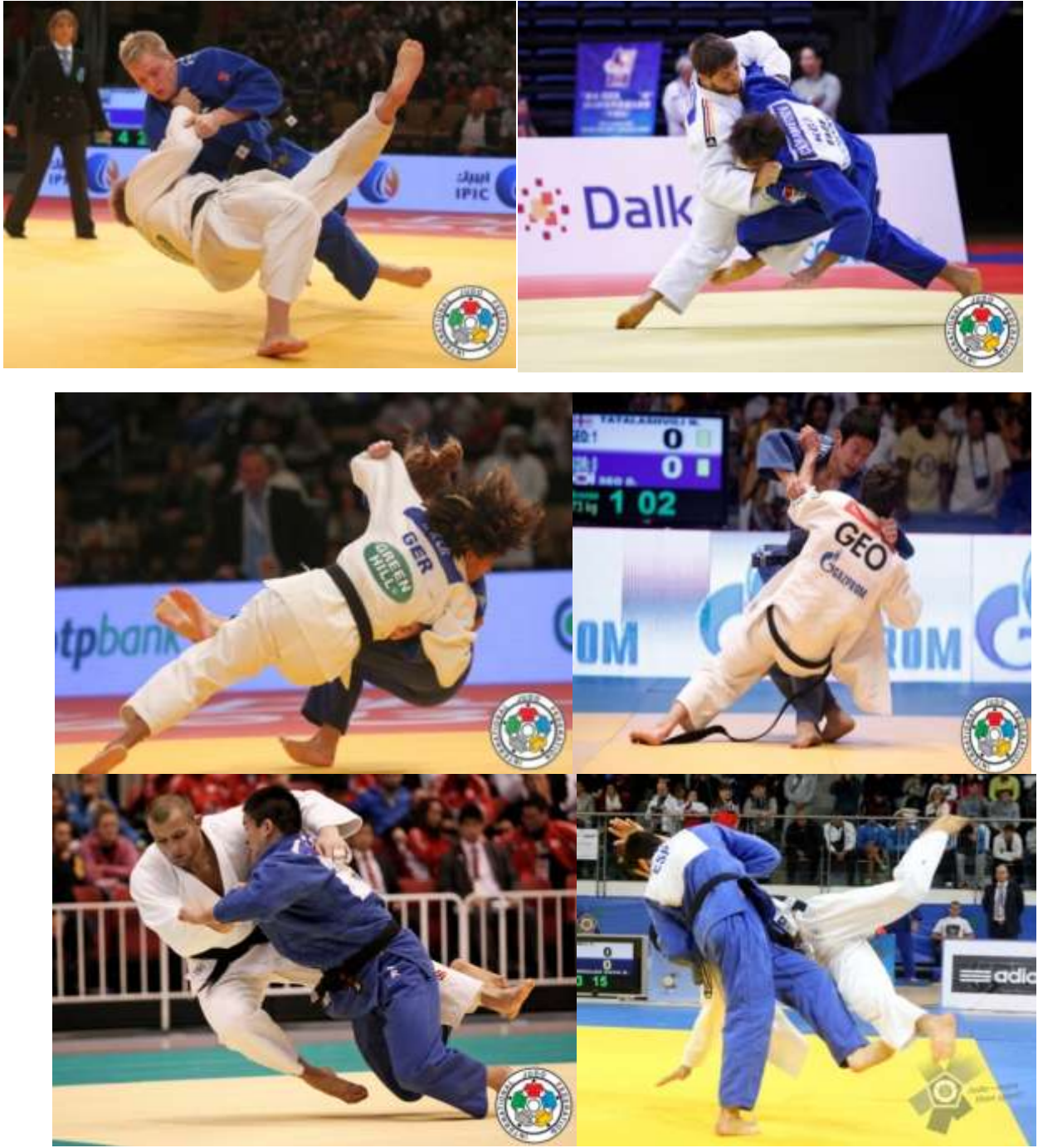

*Fig58-63 The forces application in two time ( momentum) moves the Uke's Center of Mass along a pseudo-circular trajectory increasing the effectiveness of Couple Throws with a rotational Lever, in some specific situation: De Ashi Arai and O Uchi Gari. [36]*

All that is usual in high level competitions with the application of this tool with rotation in the Transverse Plane ( horizontal) with vertical axis.
These actions could be seen as rotational application of techniques, like: Ko Uchi Gari, O Uchi Gari, Okuri Ashi, De Ashi, etc.

*2)* Sometime, in the complex dynamical situations that happen in high level competitions the tool applied can be evaluated as a dynamic solution of not wanted or prepared movements but the result of dynamic without effective "control" from Tori.
In fact the change from Couple to Lever is produced not only by a time delay but also by a change in direction of forces, because not parallel forces produce a Momentum and not a Couple.
*Fig 64-66*

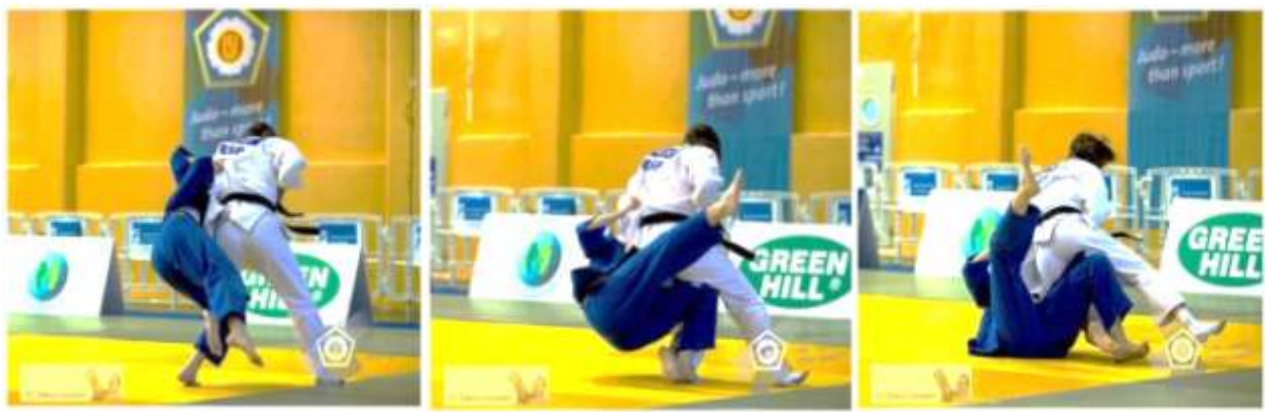

*Fig 64-66, Diagonal attack with a Couple Throw followed by a change in direction of one force, this situation change the Couple attack O Soto Gari in a Lever attack O Soto Otoshi .[36]*

## 6. New ways: Inverse applications and Chaotic variation

How is clear the Couple group forecasts simpler movements, than Lever group in other words or right closing distance or inward rotation with added trunk –leg Couple application.
This simple arrangement is the first difficulty both for coaches and athletes to find or invent new application the so called "Chaotic Variations".
In fact till now only two kind of different actions came to light the Inverse of Uchi Mata born in the Russian school , that today very utilized in high level competitions.
These applications are grounded on the Trunk-Leg Couple, but performed with inverse movement in Sagittal Plane. Fig.67

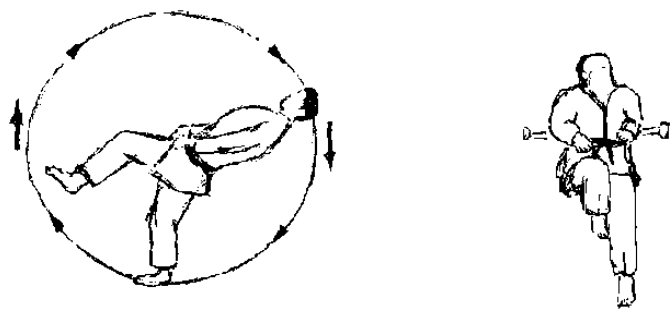

*Fig 67 Basic movement of inverse Uchi Mata and/or inverse O Soto Gari [18]*

Clearly the less mobility of spine in back direction produces a less fluid and more expensive movement.

Like classic Uchi Mata and O Soto Gari the way to close distances ( GAI) are practically two:
Right trajectory , Rotational trajectory with 180° of rotation.

Obviously the mechanics of Inverse Couple entails an external trajectory when Tori apply an Inverse O Soto Gari.

Un other interesting notation is that trajectories of contact (GAI) in these Inverse Throws are even they inverted: in Inverse O Soto Gari the sweeping leg is applied at back of leg (gleteus area) and the contact trajectory is with external rotation, the other side Inverse Uchi Mata is applied after a right contact trajectory (GAI).

This is the biomechanical explication of these Russian Throws that today are very well managed also by Japanese Athletes.

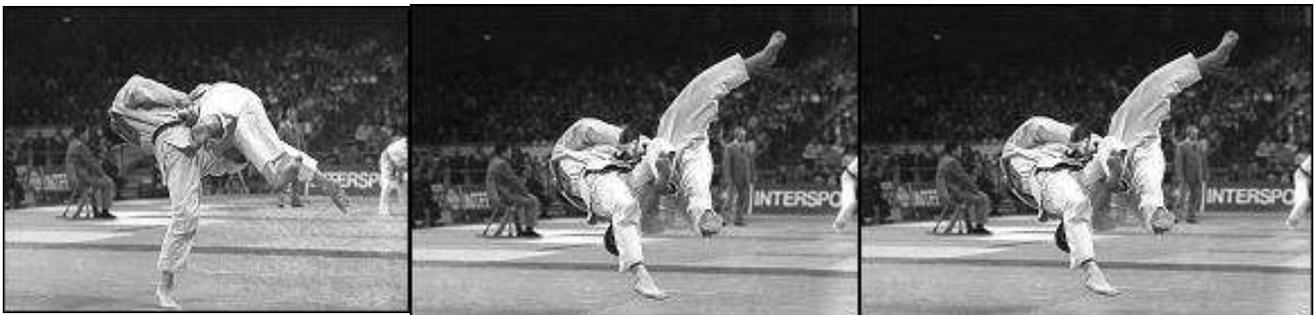

*Fig. 68-70: gyaku- (or ushiro-) Uchi Mata by Shota Chochosvili (Finch)* [18]

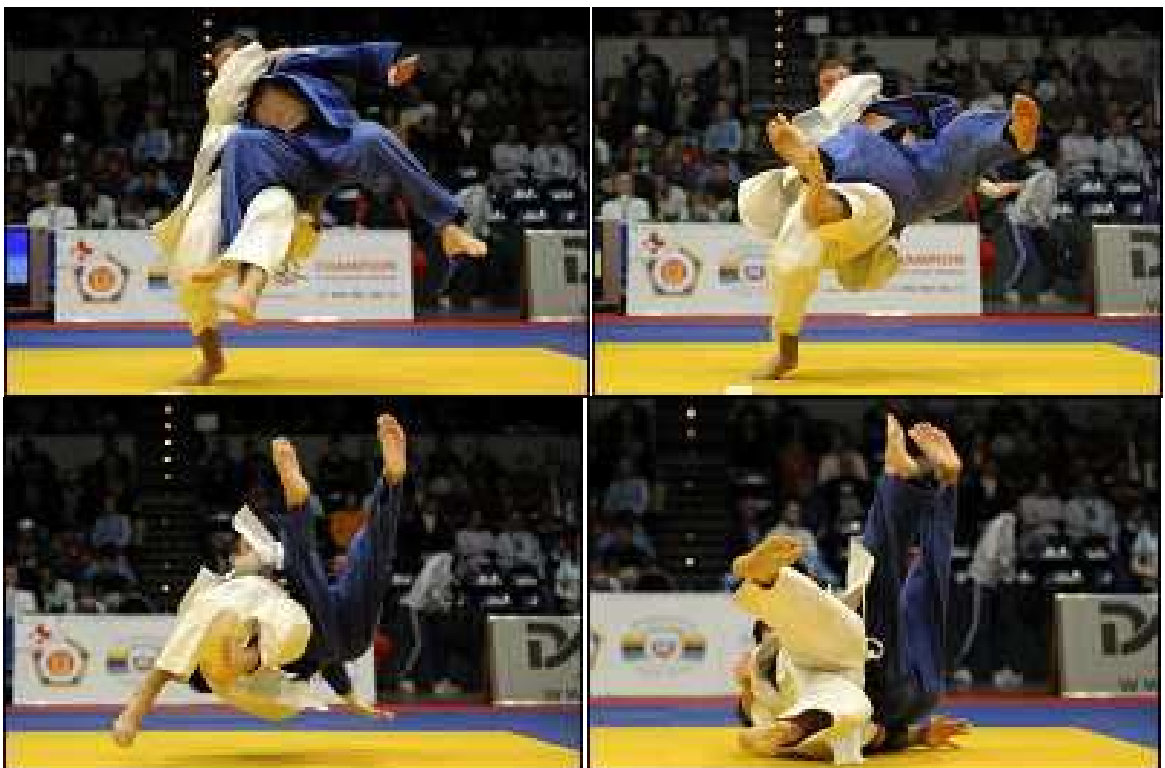

*Fig 71-74 Inverse O Soto Gari [18]*

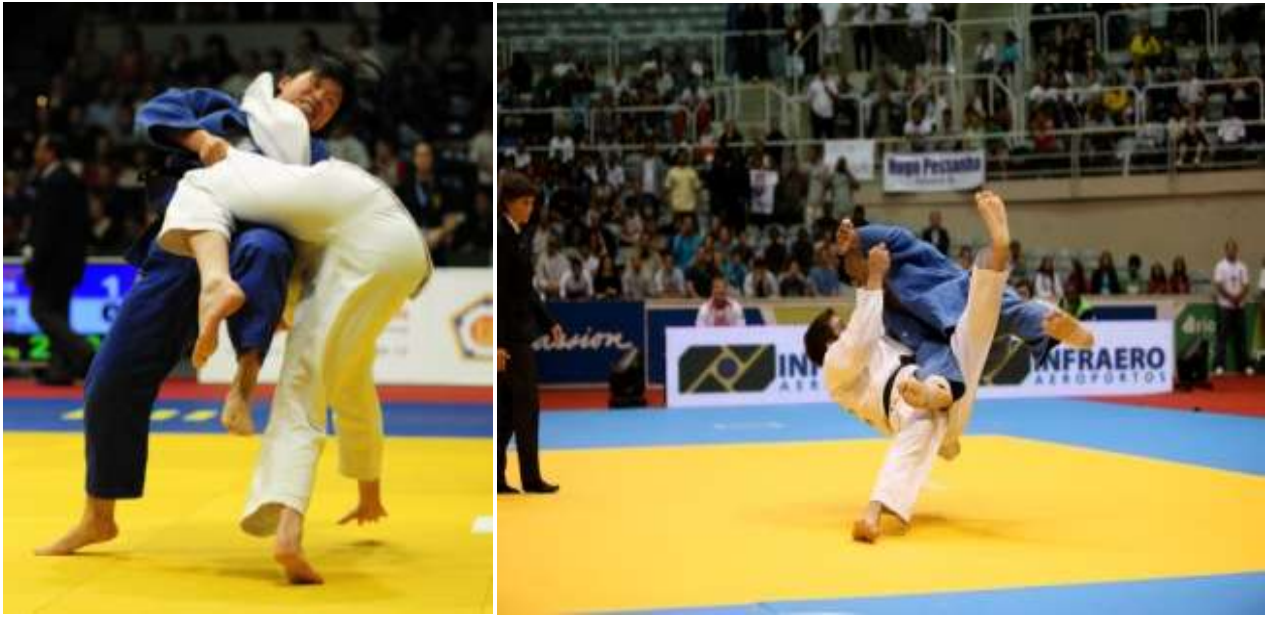

*Fig. 75-76 Inverse Uchi Mata and Inverse O Soto Gari [18]*

Different evaluation is performed for "Chaotic" variation of the Trunk-Leg Couple Group because the action movements, as already explained, are simpler than Lever actions; in effect every throw is simply ( GAI + Couple) ; it is very difficult to build new and Chaotic variations.
However also in this group during these years in high level competitions some "Chaotic "variations have been developed.
One by French Athletes, as we see in the following figures. **Fig.77-78**

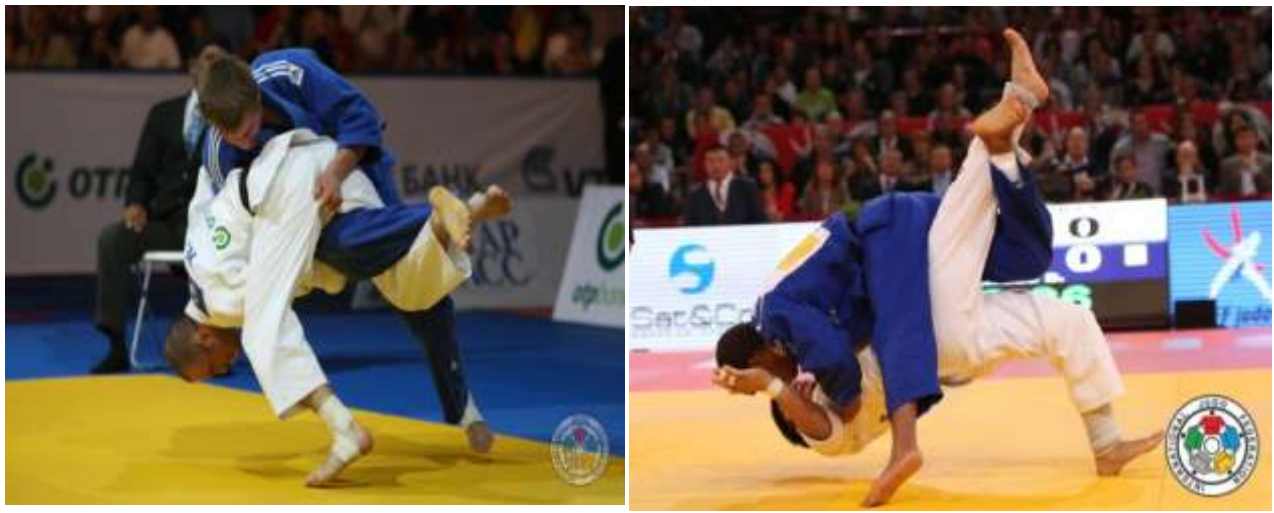

*Fig 77-78 French Chaotic Variation of Trunk-Leg Couple [21]*

Energetically very expensive, in these last times from East countries were introduced two new Trunk-Leg Chaotic Variations, grounded on one internal sweeping, always with attack front or back. **See Fig.79-80**

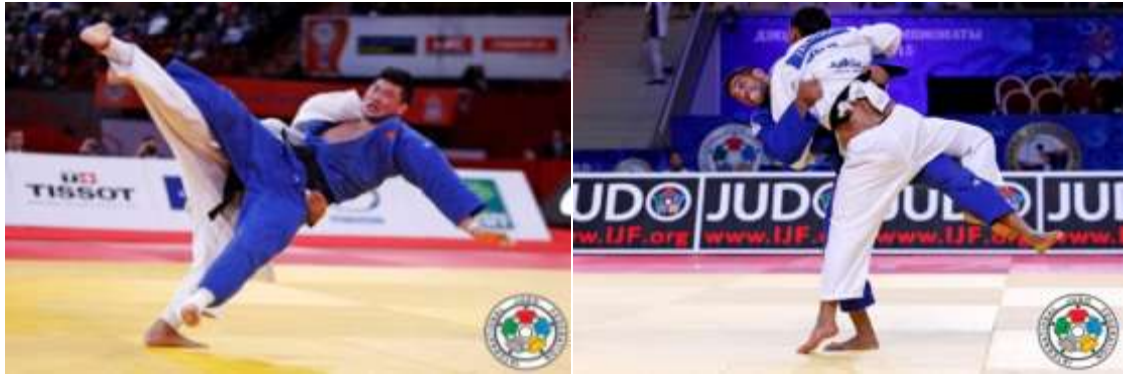
*Fig 79 80 East Countries interpretation of Trunk-Leg Couple "Chaotic Variations"[21]*

## 7. Uchi Mata Family: Physic-Biomechanical framework

Considering the Couple of Athletes, although biomechanics of Trunk-Leg Couple was analyzed in static conditions to single out the basic mechanism of throws, tanks to Galilean relativity, the same results is true also during motion in competition.

*Couple Throws*

For these throwing techniques, it is applied the principle of independent simultaneous actions, on this basis the body's complex motion in space can be simplified in sum of a motion into the Sagittal Plane plus a simple motion in space.
Then the first motion resulting by Couple applied by Tori, because is a circular plane motion is independent from gravity.
The variational analysis assures us that the first trajectory is obtained by the extremes of general function,

$$I = \int_{x1}^{x2} y^r \left(1 + y'^2\right)^{1/2} dx$$

For r=-1 with solution x=a- b sinθ and y = c – b cosθ in this case the extremal is the arc of circle with ray b and center B (a, c).

The first motion will be described by the equation (θz=s):

$$\frac{2}{3} M l^2 \ddot{\theta} = M l \frac{d^2 s}{dt^2} \quad \text{or} \quad (z - \frac{2}{3}l)\ddot{\theta} + 2\dot{z}\dot{\theta} + \ddot{z}\theta = 0$$

The second trajectory is the sum of a gravity motion plus a secondary Couple that produces a momentum in the perpendicular plane to the gravity force. This second trajectory is is more or lessa an arch of a parabola with vertex V coincident with the rotation center B or in the center of Mass of Uke if we refer at the "perfect Uchi Mata" [37]

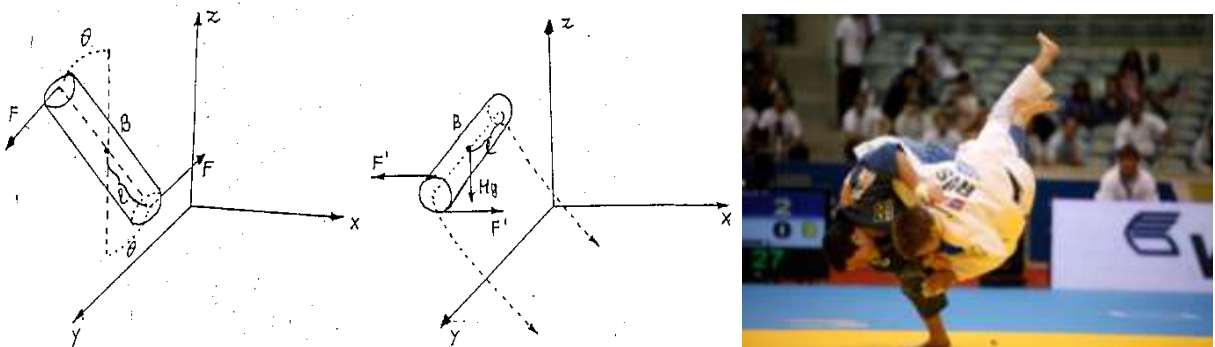
*Fig 81-83 Basic Mechanics of Trunk-Leg group throw: Uchi Mata [37]*

The application of secondary Couple in classic throw is important to achieve referee result, and it is essential in inverse throws because without Tori can go in disadvantage.

*Interation.*

How already explained in previous paper the right biomechanical way to study judo competition, is to divide it in two fields of analysis: motion of Couple of Athletes system and Interaction between athletes ( Throwing techniques ).[35]
In this paper the study of interaction is specialized only at Trunk-Leg Couple throwing Techniques.

*Grips and Trajectories (rectilinear or inward rotation)*

It is interesting to underline that grips for trunk-leg Couple group have a value lower than in Lever techniques ( es. Seoi).

Really speaking in this situation their main function i sto connect Uke 's body to its own trunk , because in such way superior force is more effective. Central double grip can help lift and Couple application in the Frontal plane ( es. Okuri Ashi Arai), while in the case of straight attack, grips will have function of shortening distance between atletes. ( es. O Soto Gari, O Tsubushi).

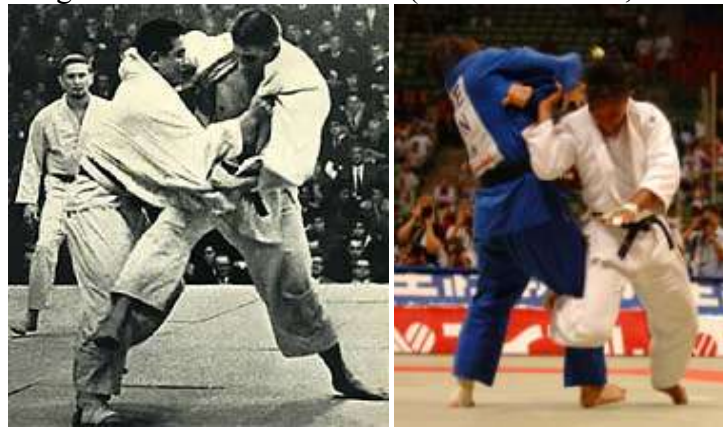

*Fig 84-85 O Sorto Gari, O Tsubushi*

Grips, during throw with inward rotation, apply also a Counterclockwise Couple (or vice versa) so as to adapt Uke's trunk to the general motion.

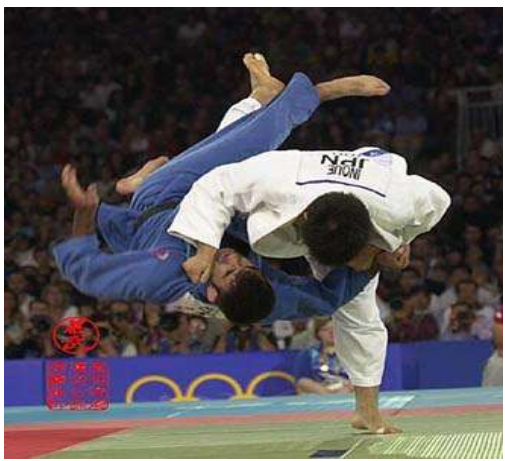 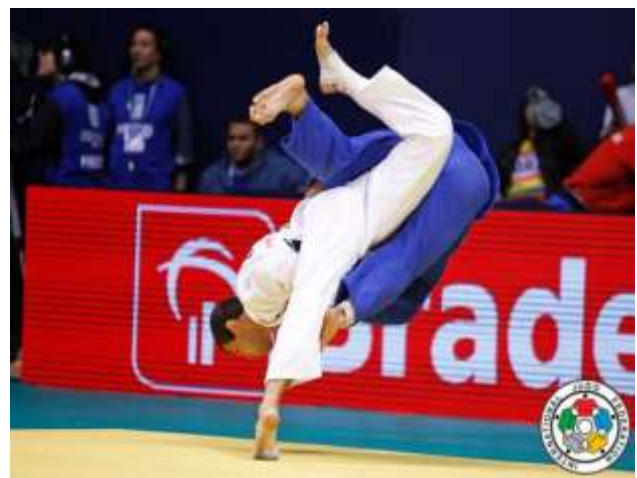

*Fig 86-87 Grips action in Uchi Mata*

As normal rule the opening of adversaries' grips is based on the movement of downward pressure followed by an instantaneous impulse upwards and a simultaneous sliding within the opponents grips .

*Lift-up*
 The action of lifting is able to facilitate throwing action by reducing the friction between the feet and Tatami, it is of great help both in the techniques of Couple in the frontal plane and with inward rotation, practically useless for Couple techniques with straight approach ( es. O Soto gari, ecc.).

*Almost plastic collision of extended soft bodies*
In the case of Group Trunk Leg both straight attack ( O Soto Gari ) and inward rotation ( Harai Goshi , Uchi Mata etc . ) Surface contact collision greater depends upon the chest that must apply upper force of Couple to opponent's chest , collision of a smaller area is the responsibility of the leg that applies lower Couple force to the opponents' leg.
So in the case of Couple at the end of the previous trajectories occurs a simultaneous contact/collision that can be considered almost plastic , because the two athletes are closely related. The mechanics is obviously different from the Lever, in this case Uke subjected to the Couple tends to rotate around its center of mass, and more is the simultaneous application of the forces and longer equivalent their intensity, more the actual motion will approach the theoretical one's . [39]

*Fig.88-93*

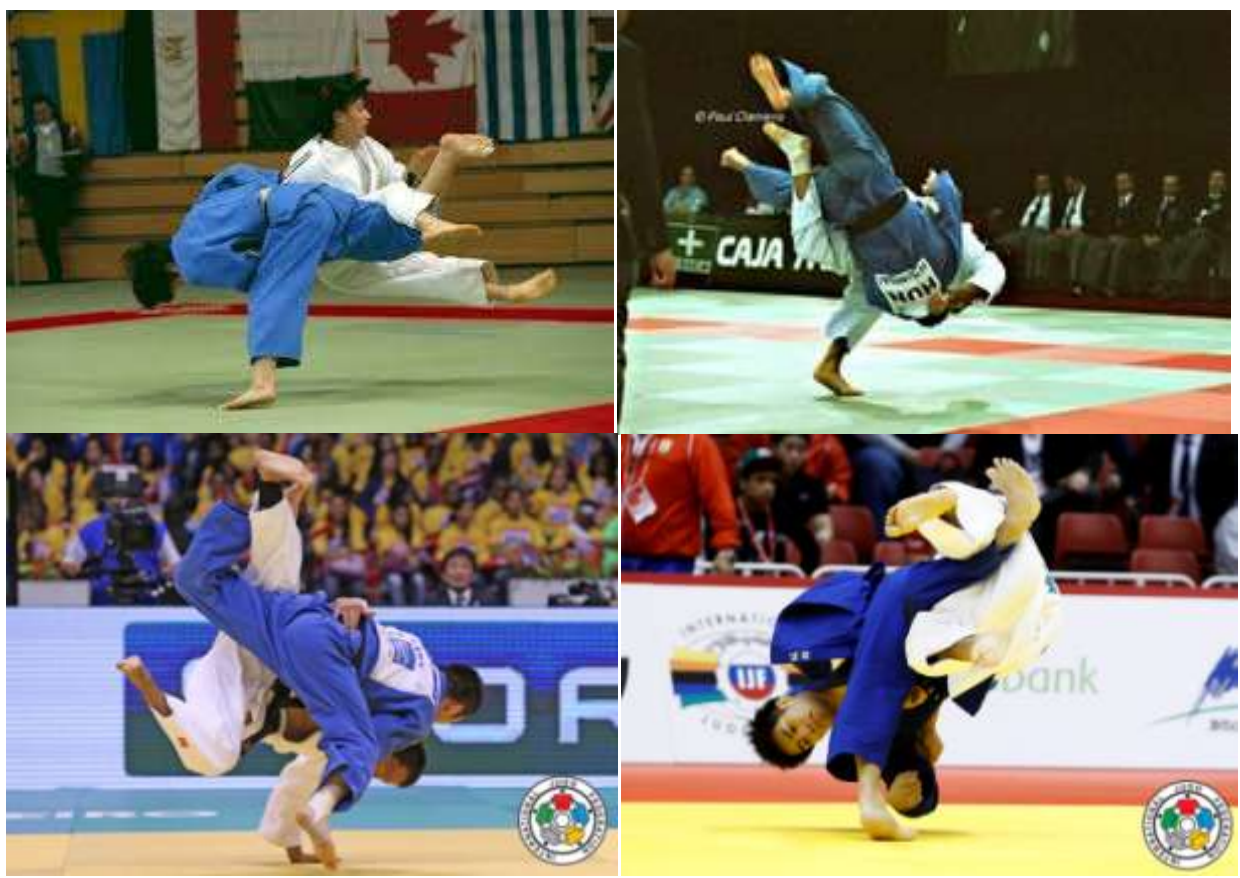

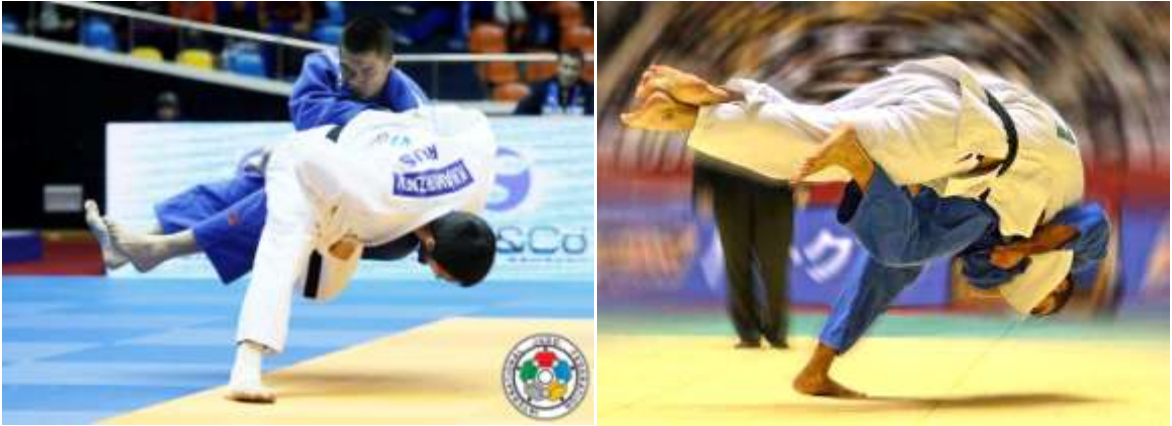

***Fig. 88-93*** *the increasing simultaneous application of the Couple approaches more and more practical technique to the theoretical ones.*

## 7. Conclusions

Biomechanics Group Couple ( trunk – leg) showed that this group of techniques that are considered among the most effective in high level competition , are in fact also energetically less expensive than those of the Lever and are also bio-mechanically simpler , relying only on closing the distance and Couple application.( GAI + Coppia).

Their intrinsic simplicity, however, hides a different complexity from those of the Lever which, remember, they need a high motor coordination to be effective.

In fact, being able to the techniques of Couple, carried out at whatever speed of displacement of Uke, require a particular timing ability.

Therefore, their specific qualities are:

1. *Increased simplicity of the technique.*
2. *Almost monotony in the technical movements.*
3. *Unbalance preferable but not essential.*
4. *Independence from Kuzushi .*
5. *Often Tsukuri easier than in Lever.*
6. *Energetically more favorable than Lever.*
7. *Specific Rhythm and Timing.*
8. *Very effective in any weight class.*